\documentclass{WileyMSP-template}
\usepackage[numbers,sort&compress]{natbib}

\usepackage{bm}
\usepackage{float}
\usepackage{hyperref}
\usepackage{ragged2e}
\usepackage{footnote}
\usepackage{stfloats}
\usepackage{footnote}
\usepackage{color}
\usepackage{makecell}
\usepackage{amssymb}
\usepackage{footmisc}
\usepackage{natbib}%
\usepackage{graphicx}%
\usepackage{amsmath}
\usepackage{setspace}
\usepackage{geometry}
\begin{document}


\title{Towards real-world quantum networks: a review}

\maketitle


\author{Shi-Hai Wei $^a$},
\author{Bo Jing $^a$},
\author{Xue-Ying Zhang},
\author{Jin-Yu Liao},
\author{Chen-Zhi Yuan},
\author{Bo-Yu Fan},
\author{Chen Lyu},
\author{Dian-Li Zhou},
\author{You Wang},
\author{Guang-Wei Deng},
\author{Hai-Zhi Song},
\author{Daniel Oblak*},
\author{Guang-Can Guo},
\author{Qiang Zhou*}



\begin{affiliations}
S. Wei, Dr. B. Jing, X. Zhang, J. Liao, Dr. C. Yuan, Dr. B. Fan, Dr. C. Lyu, Dr. D. Zhou, Prof. Y. Wang, Prof. G. Deng, Prof. H. Song, Prof. G. Guo, Prof. Q. Zhou\\
Institute of Fundamental and Frontier Sciences \& Yangtze Delta Region Institute (Huzhou) \& School of Optoelectronic Science and Engineering, University of Electronic Science and Technology of China, Chengdu 610054, China\\
Email Address: zhouqiang@uestc.edu.cn

Prof. G. Deng, Prof. G. Guo, Prof. Q. Zhou\\
CAS Key Laboratory of Quantum Information, University of Science and Technology of China, Hefei 230026, China\\

Prof. Y. Wang, Prof. H. Song\\
Southwest Institute of Technical Physics, Chengdu 610041, China\\

Prof. D. Oblak\\
Institute for Quantum Science and Technology, and Department of Physics \& Astronomy, University of Calgary, Calgary, Alberta T2N IN4, Canada\\
Email Address: doblak@ucalgary.ca

$^a$These authors contributed equally to this work

\end{affiliations}



\begin{abstract}
\justifying
\setlength{\parindent}{0em}
Quantum networks play an extremely important role in quantum information science, with application to quantum communication, computation, metrology and fundamental tests. One of the key challenges for implementing a quantum network is to distribute entangled flying qubits to spatially separated nodes, at which quantum interfaces or transducers map the entanglement onto stationary qubits. The stationary qubits at the separated nodes constitute quantum memories realized in matter while the flying qubits constitute quantum channels realized in photons. Dedicated efforts around the world for more than twenty years have resulted in both major theoretical and experimental progress towards entangling quantum nodes and ultimately building a global quantum network. Here, we review the development of quantum networks and the experimental progress over the past two decades leading to the current state of the art for generating entanglement of quantum nodes based on various physical systems such as single atoms, cold atomic ensembles, trapped ions, diamonds with Nitrogen-Vacancy centers, solid-state host doped with rare-earth ions, etc. Along the way we discuss the merits and compare the potential of each of these systems towards realizing a quantum network.

\end{abstract}


\section{Introduction}
\justifying
 Bringing together quantum physics and mathematical theory of information science\cite{planck1900improvement,dirac1930principles,shannon1948mathematical}, quantum information science (QIS) shows great potentials in today’s rapidly evolving information society, with applications ranging from advanced information collection to processing and transmission\cite{degen2017quantum,gisin2007quantum,pirandola2015advances,feynman1999simulating,knill2001scheme,kok2007linear,ladd2010quantum,harrow2017quantum,xu2020secure,OIDA:,deng2003two,long2010quantum,pirandola2020advances}. Compared to classical bits encoded as ‘0’ \textsl{or} ‘1’, QIS deals with quantum bits, which can encode ‘0’ \textsl{and} ‘1’ simultaneously, by making use of the ability to create quantum superposition. Such quantum bits, more commonly referred to as qubits, can be prepared in many physical systems, ranging from optical degrees of freedom of a single-photons, electronic energy states of (artificial) atoms, spin states of electrons or atomic nuclei, electrodynamical states of a superconducting circuits, etc.\cite{bennett2000quantum,bouwmeester2000physics}. The ability to manipulate these qubits via gate operations is at the heart of quantum computers and, although universal quantum computers or simulators have yet to be fully realized, efforts to link them together to build a global quantum network are also being vigorously pursued. The inspiration for this effort comes from the great success of the Internet, which by facilitating the sharing of classical information between computers and devices, has transformed our society. Hence, a quantum internet over which quantum information is shared between processing nodes or detectors is one of the holy grails of research in QIS. It is worth noting that there are mainly two kinds of understandings for quantum networks: one is quantum network without entangled quantum nodes\cite{xu2020secure,gisin2002quantum,frohlich2013quantum,pirandola2020advances}; the other one is quantum network with entangled quantum nodes\cite{kimble2008quantum,wehner2018quantum,simon2017towards}.

{Quantum key distribution (QKD) network is one of the typical quantum networks without the needs of entangling quantum nodes. The key goal of QKD is to achieve a communication with information-theoretical security based on the fundamental laws of quantum physics\cite{xu2020secure,zhang2018large}. Since the BB84 protocol was proposed by C. Bennett and G. Brassard in 1984\cite{ch1984quantum}, 
more and more protocols have been proposed and developed, which can be mainly divided into two categories, i.e., discrete-variable (DV) QKD and continuous-variable (CV) QKD\cite{grosshans2002continuous,grosshans2003quantum}. To solve some issues like gap between the theory and the practical implementations (due to the imperfection of devices)\cite{lo2014secure}, large-scale expanding and improvement of secure key rate, etc., there are many protocols proposed\cite{scarani2004quantum,acin2007device,hwang2003quantum,lo2005decoy,braunstein2012side,lo2012measurement,ottaviani2019modular}. Specially, decoy-state QKD\cite{lo2005decoy} and measurement-device-independent (MDI) QKD\cite{biham1996quantum,inamori2002security,braunstein2012side,lo2012measurement} have been widely demonstrated in field situations. Relying on these protocols, a number of QKD networks have been built around the world\cite{elliott2005current,peev2009secoqc,sasaki2011field,stucki2011long,mirza2010realizing,chen2009field,chen2010metropolitan,wang2014field,qiu2014quantum}, including DARPA quantum network\cite{elliott2005current}, SECOQC quantum network\cite{peev2009secoqc}, Tokyo quantum network\cite{sasaki2011field}, Hefei real-life application QKD network\cite{chen2009field,chen2010metropolitan}, and the world’s longest quantum secure communication backbone network over 2000 km, from Beijing to
Shanghai\cite{qiu2014quantum}, etc. And in 2016, QKD based on decay-state protocol was demonstrated from Micius satellite to ground over a distance of 1200 km\cite{liao2017satellite}, which proves the future possibility for building satellite based quantum network. Except for aforementioned QKD networks which are mainly applied for point-to-point communication, there are also small-scale quantum networks having been developed which suit for multi-party quantum communication. In 2020, A city-wide quantum communication network with trusted node-free eight-users was reported\cite{joshi2020trusted}. By multiplexing the spectral modes of photons, communication rate can be improved and an entanglement-based 12 wavelength-multiplexed channels quantum communication network has also been demonstrated\cite{wengerowsky2018entanglement}. Another kind of quantum network is based on entangled quantum nodes, with which qubits can be transmitted and processed directly, for instance, via quantum teleportation. In the following, we will firstly focus on such quantum network and review the recent advances of entangling quantum nodes.}

To start, in order to exchange quantum information among distant quantum nodes the consensus among researchers is to mimic the optical communication channels in classical networks and use photons to encode and carry the quantum information\cite{van1998photonic,northup2014quantum}. Therefore, the main challenge is to \textsl{faithfully transmit photonic quantum states over a suitable channel from one location to another}. Both free space and optical fibers are viable physical channels due to their low transmission loss for photonic quantum information. In free space, with the help of the Micius satellite, photonic quantum information has been transmitted over 1200 km for the application in QKD\cite{liao2017satellite} and over 1400 km for demonstration of quantum teleportation\cite{ren2017ground}. In fiber optical cable, QKD over 500 km has been reported\cite{chen2020sending,fang2020implementation}. {However, these demonstrations have approached the limits of distances achievable through direct links, which is limited by certain bounds of the optimal key rate with a point-to-point QKD protocol\cite{pirandola2009direct,takeoka2014fundamental,pirandola2017fundamental,pirandola2021limits}. More precisely, the PLOB bound\cite{pirandola2017fundamental} introduced by S. Pirandola et al. in 2015 - exactly equal to the secret key capacity of the lossy channel }- tells that the longest transmission distance with standard decoy state BB84 protocol is around 600 km with a state of the art QKD system and nearly perfect single photon detectors\cite{mao2021recent}. In classical communication, transmission loss does not limit the link distance as optical amplification – in so-called repeaters – can be used compensate the loss of the channel without degrading the quality of the optical signal. However, since quantum signals cannot be amplified (due the no-cloning theorem), the classical strategy is not suitable for the transmission of quantum information\cite{park1970concept,wootters1982single}. Hence, it is not possible to transmit photonic quantum information over arbitrary long distances directly, and the task of creating quantum links over long distances remains a significant challenge for quantum networks.

One solution to overcome the limitations set by the no-cloning theorem, was proposed in 1993 by C. H. Bennett et al. They prescribed a method for teleporting an unknown quantum state via dual classical and Einstein-Podolsky-Rosen channels, i.e. classical and quantum entanglement channels, to realize the so-called quantum teleportation\cite{bennett1993teleporting}. In 1997, quantum teleportation was experimentally demonstrated by D. Bouwmeester et al. Their experiment was based on the generation of discrete variable (DV) photonic quantum entanglement, Bell-state measurement and classical communication\cite{bouwmeester1997experimental}. This demonstration established quantum teleportation as a practical method for transmitting quantum information\cite{valivarthi2016quantum,sun2016quantum}. Thus, the task of building quantum networks includes the ability of \textsl{establishing quantum entanglement between distant quantum nodes}. It is worth noting that quantum teleportation can also be realized with continuous variable (CV) system\cite{furusawa1998unconditional,huo2018deterministic}, while in this review we focus on the quantum network based on DV systems.

\begin{figure}[ht]
	\centering
	\includegraphics[width=15cm]{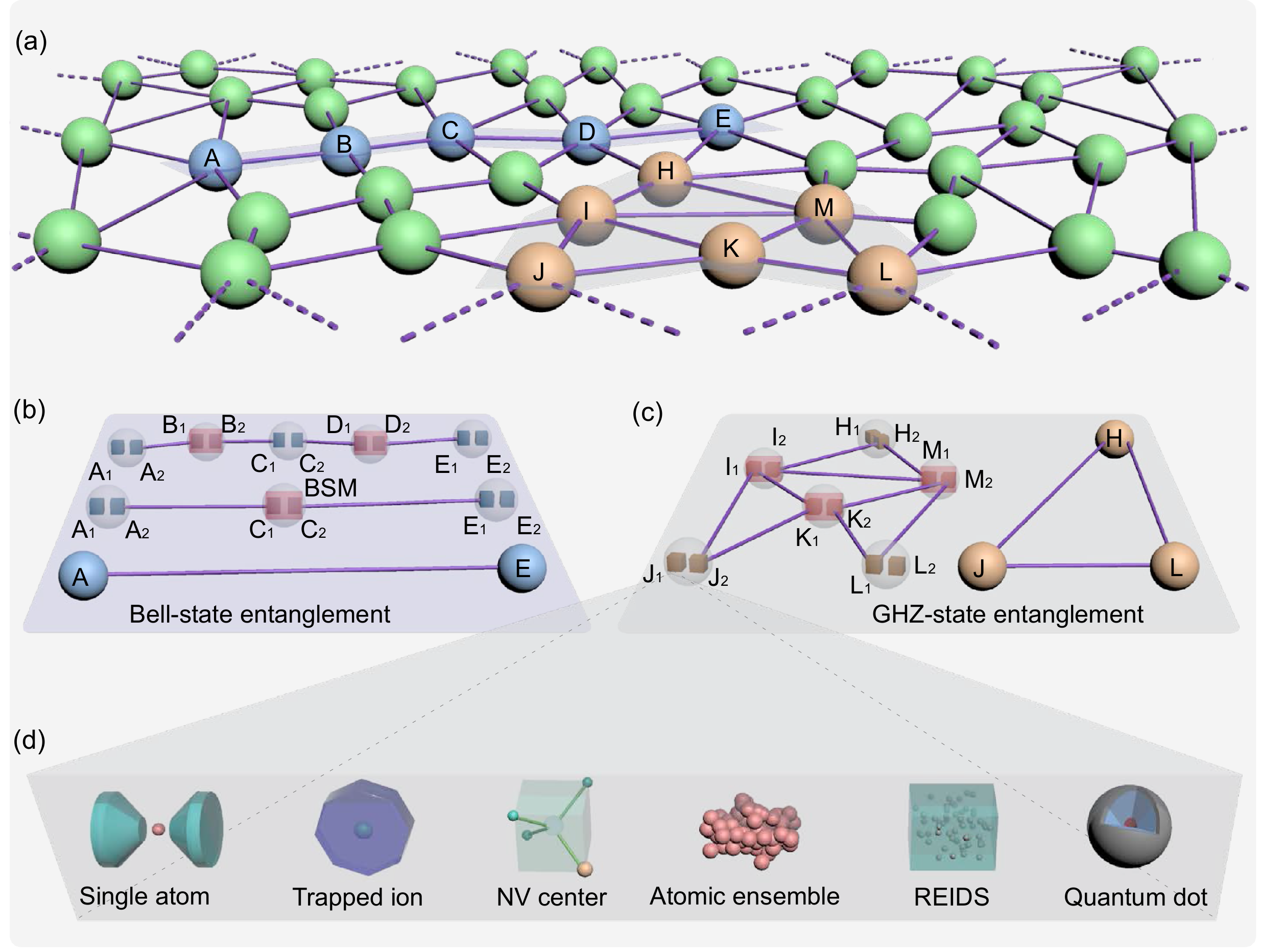}
	\caption{Typical structures of real-world quantum networks. (a) Schematic for future quantum networks with entangled quantum nodes. (b) One-dimensional quantum network. (c) Two-dimensional quantum network. (d) Typical physical platforms for quantum memories that can be placed in each quantum node. Each ball represents a quantum node while each cubic represents a quantum memory placed in the quantum node.}
	\label{fig:1}
\end{figure}

Although there is no fundamental limit to the distance over which quantum information can be teleported, the ability to distribute the required entangled state is still subject to the limit set by {the maximum rate at which two distant nodes can distribute entanglement qubits in an imperfect channel\cite{pirandola2017fundamental,pirandola2019end}}. As an alternative to direct transmission of entangled photons, in 1998, H. J. Briegel et al. proposed a quantum repeater protocol for the purpose of establishing long-distance quantum entanglement\cite{briegel1998quantum}. They suggested to divide the transmission channel into several short-distance links, known as elementary links. In each elementary link, entanglement among nodes separated by a short-distance is generated by direct photon transmission, and subsequently entanglement among increasingly farther separated nodes is realized through entanglement swapping\cite{pan1998experimental}. Since the photonic entanglement distribution over each elementary link is probabilistic, adjacent elementary links may create photonic entanglement at different times. This would mean that further entanglement swapping cannot take place, as it requires the elementary links to create entanglement simultaneously. A possible way to synchronize the generation of entanglement between adjacent elementary links, is to store the generated entanglement until entanglement has been successfully prepared in both, and then retrieve the photonic qubits simultaneously for entanglement swapping at their interface. Thus, a device for the storage of photonic quantum information is necessary for establishing entanglement of distant quantum nodes – this is the so-called quantum memory\cite{lvovsky2009optical,simon2010quantum,heshami2016quantum}. Accordingly, a final major element of building a long-distance quantum network, is the \textsl{demonstration of an elementary link in which distant quantum memories have been entangled – ideally following an entanglement swapping operation}. Once the elementary link is developed, we can build more and more elementary links and apply the entanglement swapping operation between adjacent elementary links to develop a global quantum network, as shown in Figure \ref{fig:1}.
Figure \ref{fig:1}(a) shows a schematic for future quantum network consisting of multiple entangled quantum nodes. Part of the quantum network can be built via a linear quantum network scheme with Bell states (Figure \ref{fig:1}(b)) or two-dimensional (2D) quantum network scheme with GHZ states \cite{wallnofer2016two} (Figure \ref{fig:1}(c)). The quantum nodes can be realized with various quantum memories, as shown in Figure \ref{fig:1}(d). Figure \ref{fig:1}(b) shows the generation procedures of a one-dimensional (1D) quantum network. This linear network, i.e., 1D quantum repeater, is based on Bell-state entanglement and can be used for point-to-point quantum communication between two distant quantum nodes A and E. In each quantum node, two quantum memories can be placed (denoted by the subscripts). First, entanglement is established between short-distance separated quantum memories like A$_2$ and B$_1$, B$_2$ and C$_1$, C$_2$ and D$_1$, D$_2$ and E$_1$. Second, a Bell-state measurement between B$_1$ and B$_2$, D$_1$ and D$_2$ is performed to swap the entanglement for preparing entanglement between A$_2$ and C$_1$ and between C$_2$ and E$_1$. Finally, the same operation is applied to C$_1$ and C$_2$ for generating entanglement between distant quantum node A and E. In this way, the quantum information can be transmitted from quantum node A to E, e.g., by quantum teleportation. In addition, there are also other types of quantum network that can realize quantum communication among multiple quantum nodes such as two-dimensional (2D) quantum network relying on Greenberger-Horne-Zeilinger (GHZ) entanglement\cite{wallnofer2016two}. Figure \ref{fig:1}(c) shows the generation procedures of such a 2D quantum network. The quantum states of three entangled photons (GHZ state) are mapped into three quantum memories such as I$_1$, J$_2$ and K$_1$; K$_2$, L$_1$ and M$_2$; M$_1$, H$_2$ and I$_2$, which would generate entanglement among three quantum memories and form an elementary 2D link with three memories. Following this method, more and more 2D elementary links can be built. Then the quantum swapping operation with three pairs of adjacent quantum nodes are applied on I$_1$ and I$_2$, K$_1$ and K$_2$, M$_1$ and M$_2$. Finally, a three-party entanglement channel among long-distance quantum nodes H, J and L is established. Using GHZ-state entanglement, long-distance quantum network with more nodes can be also realized. The first step is to establish GHZ-state entanglement among multiple nodes. Then more nodes are entangled through local operations, forming a small-scale quantum network. The second step is to connect multiple small-scale quantum networks and construct a large-scale quantum network\cite{pirker2018modular}.

In this review, we summarize the recent experimental progresses on the key elements needed for a quantum network, i.e., distribution of entanglement, teleportation and entanglement swapping between quantum nodes that incorporate some form of quantum memory, then we further discuss some other theoretical proposals for building a future quantum network. The review is organized as follows: Firstly, we introduced the concepts of QIS, focusing on the construction of quantum network, and a discussion of the most important steps, which is to create entanglement of distant quantum memories; Secondly, we summarize the progress of realizing global quantum network, especially entanglement of quantum memories, in different physical systems grouped according to their characteristics and physical platform; Thirdly, we discuss other potential proposals for building quantum network; Finally, we take a summarize on the review and discuss the available ways and challenges for constructing real-world quantum networks.

\section{Principles of quantum memory}
Quantum memory is the core device for building a quantum network with entangled nodes since quantum memory is able to store flying photonic qubits and keep stationary. Before reviewing the related work on quantum networks, it is useful to give a simple introduction to the realization of quantum memories. Thus, in this section, we will introduce the main proposals such as EIT (electromagnetically induced transparency) protocol, DLCZ protocol and  AFC (atomic frequency comb) protocol, all of which have been widely utilized for high performance quantum memories and entangled quantum nodes. 

\subsection{The EIT protocol}

EIT is a phenomenon that the light absorption of atoms becomes transparent or the absorption is greatly reduced under the action of control light. The atoms have $\Lambda$-type energy level structure, as shown in Figure \ref{fig:2}(a). Intuitively, the effect of control light is equivalent to opening the gap of the absorption spectrum of the atom from the middle, and extrapolating the absorption peak to both sides. The stronger the control light is, the larger gap opens. Theoretically, it can be considered that transition from ground state $|g\rangle$ to excited state $|e\rangle$ is inhibited due to the destructive interference between the probability amplitude of two transition paths, where two transition paths are $|g\rangle$ $\to$ $|e\rangle$ and $|g\rangle$ $\to$ $|e\rangle$ $\to$ $|s\rangle$ $\to$ $|e\rangle$\cite{fleischhauer2005electromagnetically}.

According to the Kramers-Kronig relationship, the abnormality of the absorption spectrum is always accompanied by the abnormality of dispersion\cite{lucarini2005kramers}. For EIT, probe light experiences a large normal dispersion process upon propagation, which leads to decreasing group velocity of light, namely the "slow light" effect. This effect can serve as a quantum memory (see Figure \ref{fig:2}(a)). The group velocity of probe light is proportional to the intensity of control light. When the probe light is transmitted in the EIT medium, the adiabatic decrease of control light can slow down the probe light. If the intensity of control light is decreased to zero, the group velocity of the probe light will decrease to zero, which means the probe light is freezed and has been stored into the EIT medium. The EIT storage progress can be described as follows: All atoms are initialized to ground level $|g\rangle$ by control light, and the atomic state can be expressed as $|\phi_{0}\rangle=|g_{1}...g_{N}\rangle$. When the probe light is stored by EIT medium, the atomic state is transferred to a collective excitation state, 
\begin{equation}
|\phi_1\rangle=\sum_{j}\phi_{j}e^{i\Delta{k}z_{j}}|g_1...s_j...g_N\rangle,
\end{equation}
where  $N$ is the total number of atoms, $\Delta{k}$ is the difference in wavevectors of the control and probe light, $z_{j}$ is the position of the $j_{th}$ atom. When the control light is turn on again, the stored  probe light is converted back and released from the atomic medium. 

\subsection{the DLCZ protocol}
\begin{figure}[ht]
	\centering
	\includegraphics[width=11cm]{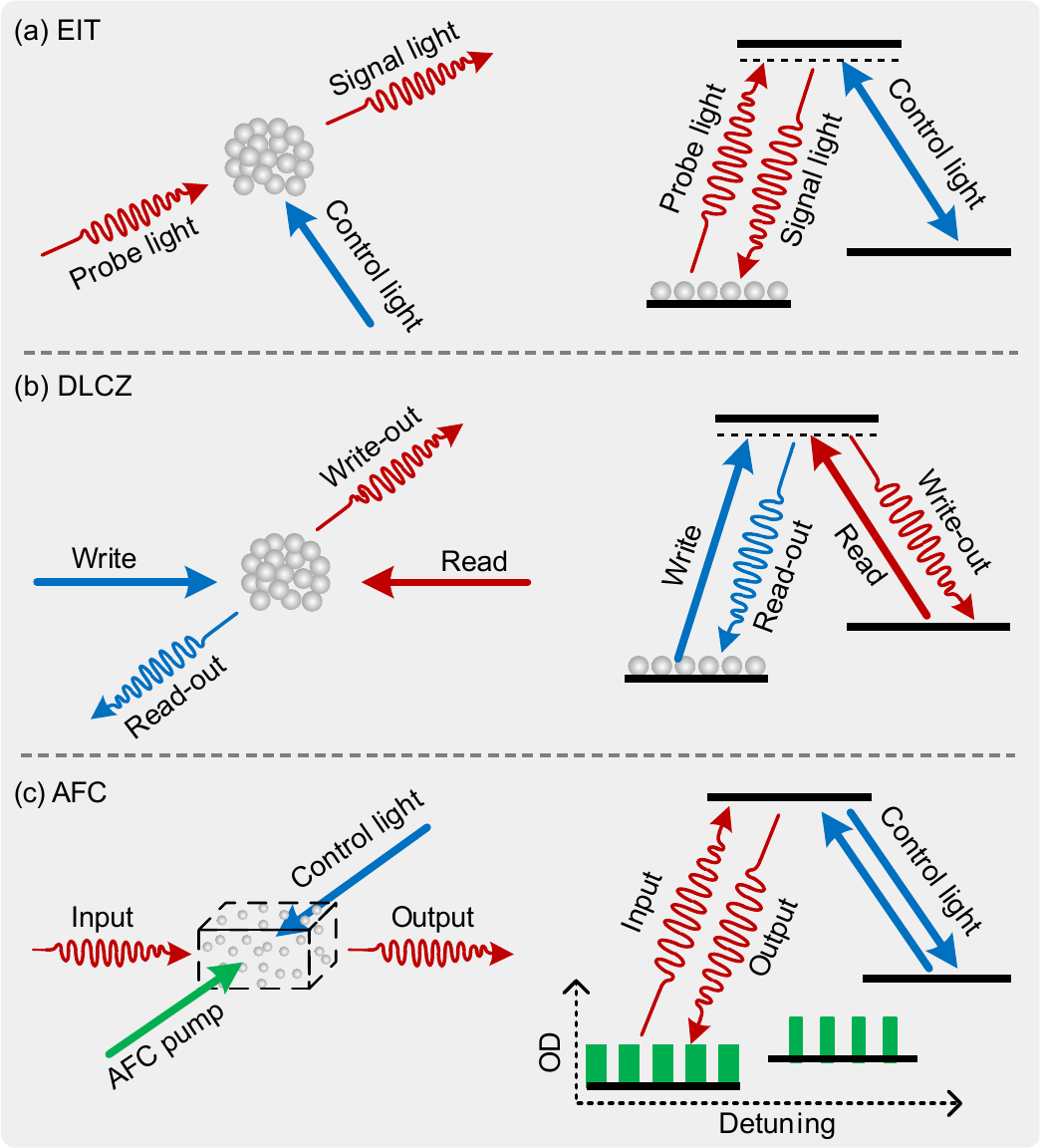}
	\caption{Protocols for quantum memories. (a) The EIT protocol; (b) The DLCZ protocol; (c) The AFC protocol.}
	\label{fig:2}
\end{figure}

{There is a quantum memory scheme similar to EIT, which is proposed by Duan, Lukin, Cirac and Zoller, i.e., DLCZ-type quantum memory protocol (see Figure \ref{fig:2}(b))\cite{duan2001long}. Compared with EIT memory, DLCZ-type memory, instead of direct storage of photonic qubit, it generates a stored qubit through a write process conditioned on emitting a photon, namely the 'write-out photon'. In this proposal, non-classical correlation between a photon and a collective atomic state is generated through a spontaneous Raman scattering process. Details for storage are described below: We also consider atoms with $\Lambda$-type energy level structure. The atomic ensemble containing $N$ atoms are initially prepared in the ground state $|g\rangle$, then illuminated by a short, weak, coherent, off-resonant write pulse with wave vector $\bm{k_w}$. A Raman transition occurs with low probability which brings one atom positioned $\bm{r_i}$ to state $|s\rangle$ while emitting a write-out photon with wave vector $\bm{k_{wo}}$. The joint atomic and photonic state can be described as,
\begin{equation}
|\psi\rangle_{wo}=[1+\sqrt{p}S^\dagger a_{wo}^\dagger+o(p)]|vac\rangle,
\end{equation}
where $S^\dagger=\frac{1}{\sqrt{N}}\sum_i^Ne^{i(\bm{k_w-k_{wo}})\cdot \bm{r_i}}|s\rangle_i\langle g|_i$ and $a^\dagger_{wo}$ are the creation operators for one collective atomic excitation and write-out photon, respectively. $|vac\rangle=|g,...,g\rangle|0\rangle_{wo}$ represents initial state, subscript $i$ represents the $i_{th}$ atom and $p$ is the probability to excite one atom, also named as excitation probability. Although the initial detection of write-out photon is probabilistic, once it has been, it will result in the conditional state with one collective atomic excitation (also called as spin-wave) $\psi_{ca}=\frac{1}{\sqrt{N}}\sum_i^Ne^{i(\bm{k_w-k_{wo}})\cdot \bm{r_i}}|g,g...,s_i,...g\rangle$, which serves as an embedded quantum memory. This excitation can subsequently be converted into a read-out photon ($\bm{k_{ro}}$) on demand with a strong read pulse ($\bm{k_r}$) and the atomic state will go back to its initial state. Similar to the write process, the emitted read-out photon state is,
\begin{equation}
|\psi\rangle_{ro}=\dfrac{1}{\sqrt{N}}\sum_i^Ne^{i\varphi} a^\dagger_{ro}|0\rangle_{ro},
\end{equation}
where $\varphi={(\bm{k_w}-\bm{k_{wo}})\cdot \bm{r_i}+(\bm{{k_r}}-\bm{k_{ro}})\cdot \bm{r'_{i}}}$ and $\bm{r'_i}$ is atom position before reading out. Now it is clear that once a phase match condition $\varphi=0$ is satisfied, the probability of the read-out photon, being emitted in the direction $\bm{k_{ro}}$, is enhanced by $|\frac{1}{\sqrt{N}}\sum_i^Ne^{i\varphi}|^2=N$ depending on the atom number $N$, i.e., the collective enhancement in CAEs. It is worth noting that the whole read process is the same as that in EIT protocol.}

\subsection{The AFC protocol}

{Atomic frequency combs (AFC) is a quantum memory protocol relying on the interaction between input photons and a medium with a comb-shaped spectrum (see Figure \ref{fig:2}(c))\cite{de2008solid,afzelius2009multimode}. In this protocol, due to rare-earth ions doped solids (REIDS) with a large inhomogeneous broadening $\Gamma_{inh}$ and narrow homogeneous linewidth $\Gamma_{h}$ in transition from ground state $|g\rangle$ to excited state $|e\rangle$ ($\Gamma_{inh}\gg\Gamma_{h}$), the comb-shaped spectrum can be produced by spectral hole burning (SHB). Details speaking, relying on spectral tailoring of inhomogeneous broadening with frequency-selective optical pump, a series of absorption peaks with a periodicity of $\Delta$ in the absorption profile of inhomogeneous broadening from the ground state $\vert$g$\rangle$ to the excited state $\vert$e$\rangle$ are produced, which is so-called AFC, and the resonant atoms are pumped to the long-lived auxiliary energy level $\vert$aux$\rangle$. When a photon resonant with transition from $|g\rangle$ to $|e\rangle$ is sent into the AFC and absorbed by the comb, a collective atomic excitation is prepared, which can be expressed as:
\begin{equation}
|\Psi\rangle=\sum_{j}^Nc_{j}e^{i2\pi\delta_{j}t}e^{ikz_{j}}|g_{1}g_{2}...e_{j}...g_{N}\rangle,
\end{equation}
where $N$ is the total number of atoms in the AFC, $|g_{j}\rangle$ and $|e_{j}\rangle$ represent the ground state and excited state of the $j_{th}$ atom, $k$ is the wave number of input light, $\delta_{j}$ is the detuning of laser frequency from atomic transition frequency, $z_{j}$ is the position of the $j_{th}$ atom, and the amplitude $c_{j}$ is determined by the frequency detuning and position of the $j_{th}$ atom. After the photon is absorbed by the AFC, the different terms in the collective excitation state $|\Psi\rangle$, having different detunings $\delta_j$, begin to accumulate different phases. Fortunately, due to the periodic structure of AFC, i.e., $\delta_{j}=m_{j}\Delta$, at time $t=1/\Delta$ all terms in the state acquire phases equal to an integer multiple of 2$\pi$, which are all equivalent to 0. This process of rephasing, leads to re-emission of the input photon at the time $t=1/\Delta$. For this AFC-based two-level storage scheme, the storage time is predetermined. Remarkably, it is also possible to perform on-demand read-out with AFC based quantum memory where a third spin state $|s\rangle$ is utilized to convert the $|g\rangle$-$|e\rangle$ coherence to $|g\rangle$-$|s\rangle$ coherence. After the input light is absorbed and stored as a collective excitation, a $\pi$ pulse of control field can be applied to transfer atoms in $|e\rangle$ to $|s\rangle$, the collective excitation is thus stored into a long-lived spin state $|e\rangle$ with the dephasing progress freezed. Finally after a time of $T$, the control field is used again to transfer the coherence back and the phase evolution continues. When the total time after absorption reaches $T+1/\Delta$, the stored photon is retrieved due to the rephasing of the collective state. A total storage time is $T+1/\Delta$ in this AFC-based spin-wave storage, where the storage time can be controlled by the spin storage time $T$.
}

\section{Quantum network with embedded quantum memories}
It is a long-standing experimental goal to connect quantum nodes and then to build a quantum network. This has seeded many groups around the world and a multitude of distinct approaches. The holy-grail of a long-distance quantum network may not yet have been claimed by any approaches, but several efforts are coming exceedingly close. Many physical systems such as single neutral atoms, cold atomic ensembles, trapped ions, and Nitrogen-Vacancy center in diamonds are not only available for direct absorption and storage of incident photons, for instance, with EIT protocol, but also advantageous over directly generating photon-matter quantum correlations, i.e., they can act as embedded quantum memories, which is the fundamental building block and used more often for constructing quantum networks. Here, we will focus on the construction of quantum networks with these embedded quantum memories.

\subsection{Quantum network with single neutral atoms}
There are two advantages in using single atoms as quantum nodes: Quantum information mapped onto individually addressable single atoms can be controlled and manipulated easily; Single photons can control the properties of single atoms if the light-matter coupling is strong\cite{reiserer2015cavity}. Usually, single atoms are placed in optical cavities to increase the light-matter interaction strength\cite{mckeever2003state,thompson2013coupling}. Single atoms trapped in optical cavities have been widely studied in past twenty years and exhibited excellent properties especially for storage of quantum information with high fidelity and long coherence time\cite{specht2011single,rosenfeld2011coherence}, light-matter interfaces\cite{wilk2007single} and light-matter entanglement over long distances\cite{rosenfeld2008towards}, etc. Thanks to these excellent properties, the single atom system is a promise candidate for implementing the stationary quantum nodes in a quantum network.

In 1997, J. I. Cirac et al. described a protocol for quantum state transfer and entanglement distribution by utilizing strong coupling between a high-finesse optical cavity and atoms\cite{cirac1997quantum}. In this scheme, the quantum node consists of an atom coupled to the optical mode of a cavity. The internal state of the atom is mapped onto a photon wave-packet at the transmitter node by applying an strong optical control field, whereafter the wave-packet is deterministically absorbed by the atom in the cavity at the receiver node. This means that the state of photon wave-packet is transferred to the receiver atom. By doing so, two distant nodes, i.e., atoms inside the cavity, would be connected and, thus, an elementary link formed. In 2003, L. M. Duan et al. described a scheme to build entanglement between two atoms coupled to cavities\cite{duan2003efficient}. In this scheme, two atoms are placed into two separate cavities acting as quantum nodes, at the middle of the two quantum nodes, a Bell-state measurement system is inserted. In each node, the atoms have double-$\Lambda$ level structure, where the states $|g\rangle$, $|0\rangle$ and $|1\rangle$ are three hyperfine energy levels of ground state, and $|e\rangle$ is the excited state. The initial atomic state is prepared to the the ground state $|g\rangle$. A classical laser pulse adiabatically drives the atom to other ground states ($|0\rangle$ or $|1\rangle$) through Raman process. Depending on different polarizations (H/V) of the emitted photon, the atom would end up in different Zeeman-splitting state ($|0\rangle$ or $|1\rangle$), as a result, entangled state between photon and atom is created as,
\begin{equation}
|\psi\rangle_{atom-photon}=\frac{1}{\sqrt{2}}(|0,H\rangle+|1,V\rangle).
\end{equation}
Bell-state measurements are performed on two photons from two nodes. According to the proper measured results, the maximal entanglement between two atoms coupled in the two cavities (marked as $L$ and $R$) are generated and can be prepared as,
\begin{equation}
|\psi\rangle_{atom-atom}=\frac{1}{\sqrt{2}}(|0,1\rangle_{LR}+|1,0\rangle_{LR}).
\end{equation}

\begin{figure}[t]
\centering
\includegraphics[width=16cm]{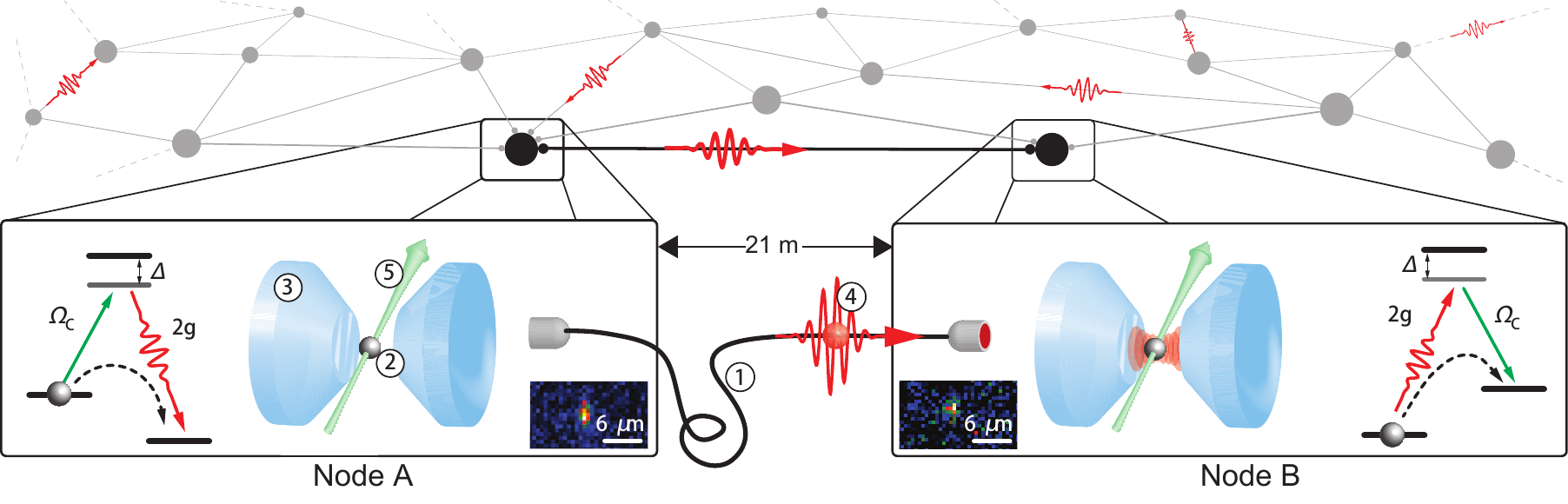}
\caption{An elementary quantum link based on single atoms in cavities. Two quantum nodes (A and B) separated by 21 m are connected through a 60 m optical fiber \textcircled{1}. In each node, a single Rb atom \textcircled{2} is trapped in an optical cavity \textcircled{3}. Insets are fluorescence images of two $Rb$ atoms and atomic level scheme, where green (red) arrow represents the control laser (exchanged single photon), $\Omega_{c}$ is the Rabi frequency of the control laser field and 2$g$ is the vacuum Rabi frequency of the vacuum mode of the cavity. Quantum states would be transferred between two atoms by exchanging a single photon \textcircled{4}. Two external laser fields \textcircled{5} control coherence internal states of atoms. At node A, by applying a laser field, the internal state of the atom is entangled with the polarization state of a photon. At node B, by applying the laser field, the state of the photon is converted to the internal state of another atom. Finally, remote entanglement between two atoms is generated. Adapted from Ref. \cite{ritter2012elementary}. }
\label{fig:3}
\end{figure}

To realize entanglement among these quantum network nodes, one of the most important things is to generate entanglement between a single atom and a single photon. J. Volz et al. in 2006 observed entanglement between a photon and a single atom\cite{volz2006observation}. In 2008, W. Rosenfeld et al. sent the photon through a 300 m fiber and demonstrated entanglement between a single trapped atom and a single photon at remote locations\cite{rosenfeld2008towards}. To utilize the strong light-matter interaction, in 2007, T. Wilk et al. experimentally generated single atom-single photon entanglement inside an optical cavity\cite{wilk2007single}, and further in 2012, S. Ritter et al. reported entanglement between two single atoms\cite{ritter2012elementary}. In their experiment, the nodes, which consisted of single trapped rubidium atoms coupled to Fabry-Perot resonators were placed in two laboratories at a distance of 21 meters and connected by a 60 meter fiber link, as shown in Figure \ref{fig:3}.
First, entanglement between the photonic polarization state and the internal state of the atom was prepared through Raman scattering. Then the photon was sent to another atom through 60 meters of fiber and stored using the electromagnetically induced transparency (EIT) quantum memory protocol \cite{marangos1998electromagnetically,fleischhauer2005electromagnetically}. Through the EIT process, the polarization of the photon was mapped onto the atomic state, and finally an entangled state between two single atoms was attained,
\begin{equation}
|\psi^{-}\rangle_{atom-atom}=\frac{1}{\sqrt{2}}(|1,-1\rangle\otimes|2,-1\rangle-|1,1\rangle\otimes|2,1\rangle),
\end{equation}
where $|1,\pm1\rangle$ ($|2,\pm1\rangle$) represents $|F=1,m_{F}\pm1\rangle$ ($|F=2,m_{F}\pm1\rangle$); $F$ is the hyperfine ground state and $m_{F}$ is the Zeeman state.The entanglement fidelity was up to 98\% between the two atoms (i.e. two nodes). The successful probability of entanglement creation between two remote atoms was 2\%. This was the product of the efficiency of atom-photon entanglement generation of 40\%, the photon transmission efficiency of 34\%, and the photon storage efficiency of 14\%. The entanglement verification efficiency was $3.2\times10^{-5}$, including the intrinsic retrieval efficiency of 0.6, the detection efficiency of 0.3 and the photon mapping efficiency of 0.03. Combined, these factors finally lead to a low atom-atom entanglement generation efficiency. Accounting for an experimental repetition rate of 5 kHz, the final photon coincidence count rate was about 3 per minute. This was owning to the existence of several primary technical challenges, including optics losses, atomic localization\cite{reiserer2015cavity}, and quantum errors occurring in the process of photon transmission\cite{northup2014quantum}. Till now a single-atom based quantum memory with good quality is still a necessary component for building a quantum network with single atoms.

To increase the collection efficiency of single photons over the whole 4$\pi$ solid-angle of emission and to improve the atom-atom entanglement generation rate, apart from utilizing a cavity to enhance the photon emission probability into the cavity mode, it is also workable to use a high numerical aperture (N.A.) lens for collection of photons, which was done by J. Hofmann et al. in 2012\cite{hofmann2012heralded}. Unlike the direct storage of photons from atom-photon entanglement to obtain probabilistic atom-atom entanglement, J. Hofmann et al. realized heralded entanglement of two trapped single $^{87}Rb$ atoms separated by 20 meters via Hong-Ou-Mandel (HOM) interference between two single-photons\cite{hong1987measurement}. In their experiment, they first separately generated entanglement between the Zeeman state of two $^{87}Rb$ atom and two photonic polarization states through Raman scattering. Then the two photons were overlapped on a beam splitter and a Bell-state measurement was applied to finish the entanglement swapping operation. The Bell-state measurement results would herald the existence of entanglement between two atoms. As for any Bell-state measurement, the temporal, spatial and spectral indistinguishability of the partaking photons are the main factors affecting the fidelity of swapping operation. After optimizing the indistinguishability of the photons, atom-atom entanglement was accomplished with a fidelity of 92\%. The overall atom-atom entanglement generation efficiency was $0.54\times10^{-6}$, including an atom-photon entanglement generation efficiency of $0.9\times10^{-3}$ and $1.25\times10^{-3}$ for the two atoms, respectively. Accounting for an experimental repetition rate of 50 kHz, the whole photon coincidence count rate was about 0.6 per minute.
\begin{figure}[t]
\centering
\includegraphics[width=12cm]{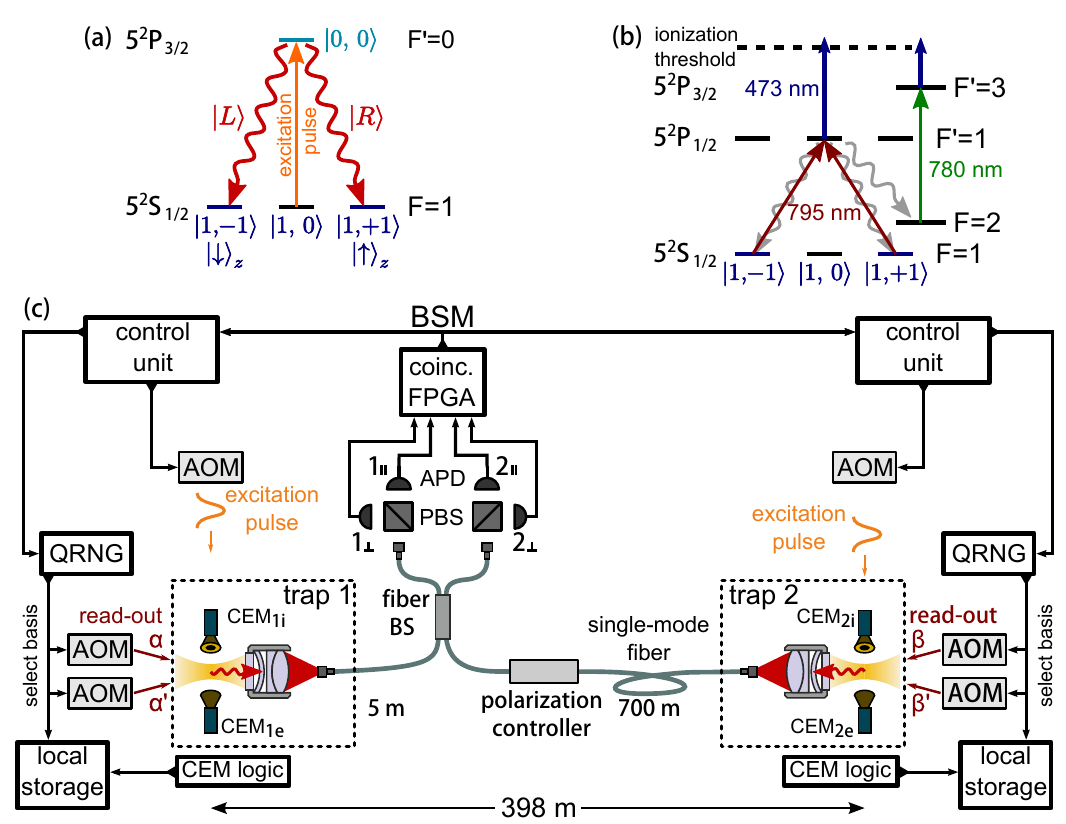}
\caption{Event-ready Bell test based on entanglement between two atoms.(a) Atomic levels for entanglement generation between the photonic polarization state and the atomic spin state. The entanglement generation is determined by the spontaneous emission of the excited state after optical excitation. (b) The measurement of atomic states. By applying a 795 nm optical pulse, the spin superposition state is excited to the state of $5^{2}P_{1/2}$, $F^{'}$=1, and ionized by applying a 473 nm optical pulse. Then the atom would decay to the level of $5^{2}P_{1/2}$, $F^{'}$=1 or $F^{'}$=2 through spontaneous emission. The measurement process would be influenced when the atom decays to $F$=1. When the atom decays to $F$=2, the population can be excited by a 780 nm laser and ionized. (c) Scheme of the experimental setup. The entanglement between spin state of the atom and polarization state of the photon is generated in each trap. The photons interfere on a 50:50 beam splitter, and the results herald the entanglement generation between the spins of two atoms. Local measurements are carried out on the spins of atoms according to settings selected by quantum random number generators. Adapted from Ref. \cite{rosenfeld2017event}.}
\label{fig:4}
\end{figure}

It is difficult to extend the entanglement of single atoms to more quantum nodes mainly because the overall photon detection efficiency is low, which leads to low atom-atom entanglement generation and verification efficiency. One contribution to the low efficiency is the verification process of the atomic state, which is usually done by retrieving the atomic state into photonic state and detecting the photons. This verification procedure will inevitably introduce non-unity retrieval efficiency, collection efficiency, transmission efficiency, detection efficiency, and so on. With more atomic states to be measured, the overall efficiency will undergo an exponential decrease, preventing connection of more quantum nodes. Fortunately, there is another way to measure the atomic state with 100\% efficiency through ionizing the atomic state. It was applied to verify the atom-atom entanglement by W. Rosenfeld et al. in 2017\cite{rosenfeld2017event}. The experimental layout is shown in Figure \ref{fig:4}. In their experiment, the atom-atom entanglement was prepared through a similar procedure to that used in Ref. \cite{hofmann2012heralded}. The complete entanglement verification efficiency was $0.7\times10^{-6}$, including atom-photon entanglement generation efficiencies of $1.65\times10^{-3}$ and $0.85\times10^{-3}$ or the two atoms, respectively. Accounting for an experimental repetition rate of 52 kHz, the whole photon coincidence count rate was about 2 per minute. With a near-unity measured efficiency of atomic state, this approach is able to close the detection loophole in the Bell-inequality test. Combined with a fast measurement, the locality loophole can also be closed. Thus, through their experiment, an event-ready Bell test based on two pairs of atom-photon entanglement was performed. The evaluation of local hidden variables theories using entangled single atoms with no locality or detection loophole was thus demonstrated.

As a summary, an approach based on single atoms is promising for quantum network, because it can directly generate high-purity atom-photon entanglement without any probability of multi-excitations, which are a crucial issue in cold atomic ensembles (See next section), and it gives access to long coherence time and strong light-matter interaction\cite{reiserer2015cavity}. However, there are still many factors limiting its further progression, mainly concentrating on single atom quantum memory. One of the factors is the time needed to load a single atom, usually on the order of ten milliseconds\cite{fortier2007deterministic}, which limits the repetition of storage. Another limitation is the efficiency of photon collection, which can be improved through cavities with ultra-low round-trip loss or a large N.A. lens for collection of photons. The third limitation is the bandwidth. The Fourier-limited bandwidth of single atom quantum memory is determined by the natural linewidth (several MHz) of the dipole transition in the atom. Compact cavity designs may decrease the lifetime of the transition through Purcell enhancement, thus, increasing the natural linewidth. While extending the distance of the nodes, photon transmission loss and memory lifetime will also dominate the overall efficiency. To solve these problems, it is necessary to use telecom-band photons for transmission. The lifetime of memory should be further improved which is on the order of 100 $\mu$s with current states of the art, limited by the fluctuation of magnetic field, residual state-dependence of the optical trapping potential and the thermal energy of the trapped atom, etc.

\subsection{Quantum network with cold atomic ensembles}
Another well-studied physical system for quantum network is cold atomic ensemble (CAE)\cite{sangouard2011quantum}. Trapping of single-atoms and realizing strong light-atom interaction and collection of a single photons requires sophisticated experimental technologies. Hence, the motivation of using CAEs as nodes is in part that less time consumed for the preparation of atoms and the experimental setups, though not simple, are more common and have been perfected over time. Furthermore, strong and controllable coupling between atoms and photons can be acquired much easier due to the collective enhancement. 
\subsubsection{Entangling quantum memories via single-photon detection}
In 2001 L. M. Duan et al. proposed a scheme based on single-photon detection to implement a quantum repeater and realize scalable long-distance quantum communication with CAEs\cite{duan2001long}. Using DLCZ quantum memory schemes, probabilistic write-out photons are generated from two remote CAEs, where write-out photons are non-classically correlated with the collective atomic states. Then the write-out photons are sent to a beam splitter (BS) for single-photon interference. It is essential that the write-out photons from both memories are indistinguishable, thus, once one and only one photon is detected by the single photon-detectors at either output ports of the BS, there is no way to distinguish which CAEs the detected photon comes from. As a result, a heralded Fock-state type entanglement between the two ensembles is generated,
\begin{equation}
|\psi\rangle_{atom-atom}=\frac{1}{\sqrt{2}}(|0\rangle_a|1\rangle_a+|1\rangle_a|0\rangle_a),
\end{equation}
where whether there is a collective state in the CAE is encoded as Fock-state qubit $|1\rangle_a$ and $|0\rangle_a$. The relative phase between $|0\rangle_a|1\rangle_a$ and $|1\rangle_a|0\rangle_a$ has been set to zero.

\begin{figure}[t]
\centering
\includegraphics[width=12cm]{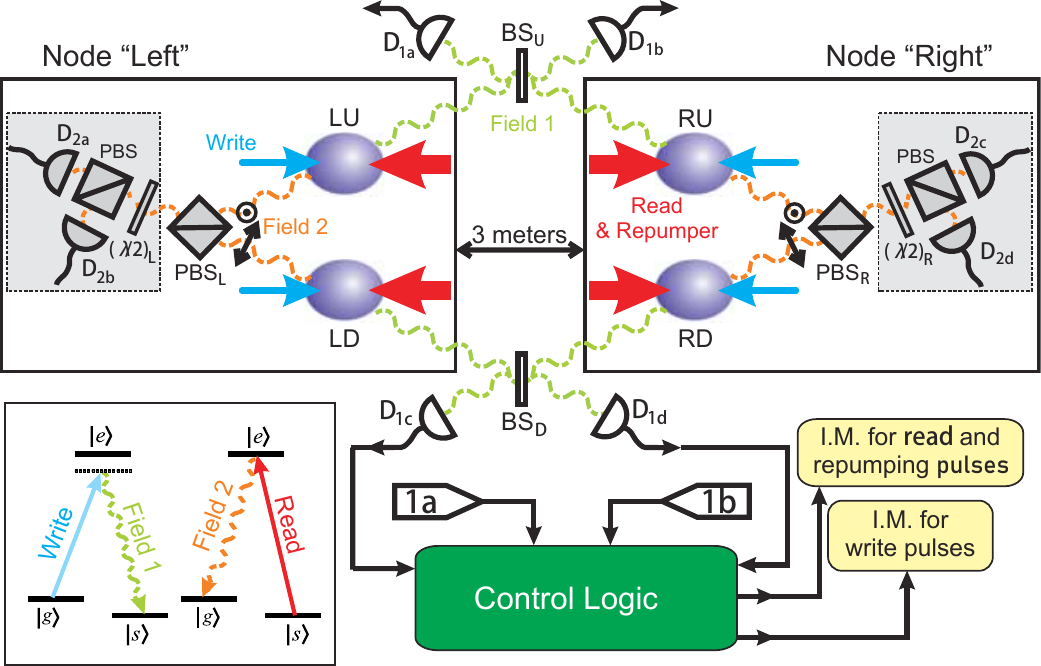}
\caption{Entanglement distribution between two quantum nodes. Entanglement between LU and RU (LD and RD) are generated while either detector $D_{1a}$ or $D_{1b}$ ($D_{1c}$ or $D_{1d}$) is clicked. After the entanglement is generated on both pairs of cesium atom ensembles, the collective excited state of atom ensembles is mapped to read-out photon by applying the read pulses. The read-out photons are combined on the polarizing beams splitter PBS$_{L}$ (PBS$_{R}$) to generate field $2_{L}$ ($2_{D}$). If coincidences are acquired between fields $2_{L}$ and $2_{D}$, the entangled state is polarization maximum entangled state. The inset is the relevant atomic levels. Adapted from Ref. \cite{chou2007functional}.}
\label{fig:5}
\end{figure}

Returning to the experimental progress in CAEs, it was 2003 that A. Kuzmich et al. and C. H. van der Wal et al. experimentally realized a quantum memory using DLCZ protocol with trapped cold caesium atoms\cite{kuzmich2003generation} and rubidium atoms\cite{van2003atomic}, respectively. Going further, to realize long-distance quantum communication with DLCZ-type quantum memories, it is essential to entangle two quantum memories\cite{duan2001long}. Based on this scheme, in 2005, C. W. Chou et al. reported the experimental entanglement generation between two caesium atomic ensembles\cite{chou2005measurement}. In their experiment, two classical write pulses synchronously illuminated two cold caesium atomic ensembles separated by 2.8 meters and write-out photons were scattered through the Raman process. Then the write-out photons were interfered on a 50:50 BS and detected by photon detectors to herald the generation of entanglement between the atomic ensembles. To verify this entanglement, the collective atomic states were converted to read-out photons and the concurrence was measured. In 2007, similar entanglement between two CAEs was generated, and a detailed quantitative characterizations for the scaling behavior of entanglement with excitation probability and storage time was performed\cite{laurat2007heralded}. In the same year, C. W. Chou et al. further reported entanglement between two nodes separated by 3 meters and verified this entanglement\cite{chou2007functional}. In their experiment, each quantum node consisted of two atomic ensembles, i.e., LU and LD in node L, RU and RD in node R, as shown in Figure \ref{fig:5}. 
Firstly, entanglement between a pair of vertically displaced atomic ensembles LU and RU (LD and RD) was created by the DLCZ protocol. Then the atomic states were mapped to read-out photons through the read-out light. The read-out photons from LU and LD (RU and RD) were combined on the $PBS_{L}$ ($PBS_{R}$). When proper coincidences between outputs of $PBS_{L}$ and $PBS_{L}$ were registered, the state was a maximally polarization entangled state. Going beyond the two node demonstrations, in 2010, K. S. Choi et al. reported entanglement of four CAEs\cite{choi2010entanglement}. In their experiment, write light pulses were split to four beams and then interacted with four CAEs simultaneously, but it generated only one collective atomic state in total, which was shared by all four CAEs (i.e., a $W$-type entanglement of four CAEs using the generation of Fock-state qubits). This is the largest size of entanglement for CAEs as quantum memories at present, however it is still difficult to entangle more CAEs. When Fock-state qubits are used to connect different quantum nodes, usually only one photon is detected each time. Although it is of great advantage for high success probability of further connection, such as for entanglement swapping, the Fock-basis encoding also necessitates that the phase is kept stable over the entire photon transmission path; something that is hard to guarantee, especially over tens and hundreds of kilometers length. And the "nothing-happened" state $|0\rangle_a$ will affect the long-distance entanglement fidelity in further entanglement connection in a quantum repeater\cite{briegel1998quantum}. These disadvantages decrease the robustness of Fock-state entanglement and limit its further application in quantum network.

\subsubsection{Entangling quantum memories via two-photon detection}
\begin{figure}[t]
\centering
\includegraphics[width=14cm]{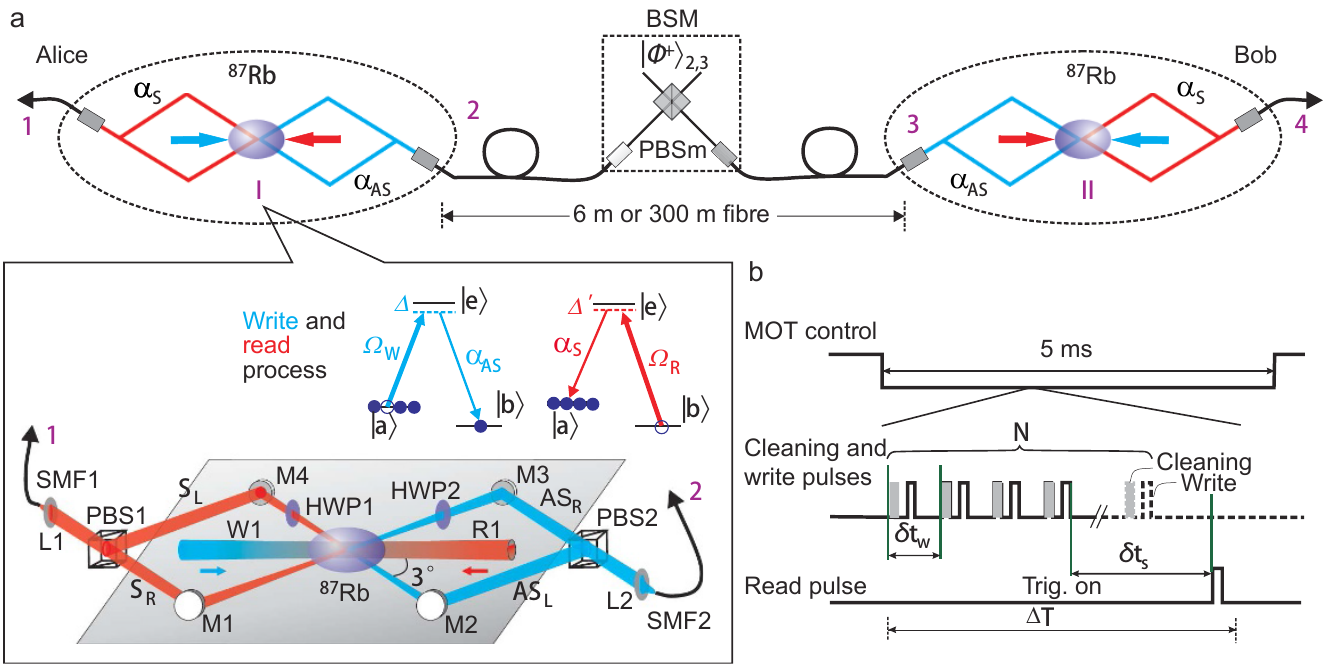}
\caption{The experimental scheme for entanglement generation based on two photon detection. (a) Photons 2 and 3 interfere in Bell-state analyzer and according to the measurement results the entanglement between two $^{87}Rb$ atomic ensembles is generated. Photons 1 and 4 (2 and 3) represent the photons from the anti-Stokes modes (Stokes modes), $|\phi\rangle_{2,3}$ is a Bell state. (b) The experimental time sequences. Adapted from Ref. \cite{yuan2008experimental}.}
\label{fig:6}
\end{figure}

To overcome the disadvantages of entanglement via single-photon detection and realize a robust generation of entanglement, two-photon detection protocols are presented, which are based on HOM interference. L. Jiang et al. presented entanglement of elementary link implemented by single-photon detection, but entanglement swapping between elementary links implemented by two-photon detection\cite{jiang2007fast}. Further, B. Zhao et al. and Z. B. Chen et al. showed entanglement of an elementary link and entanglement swapping between elementary links, which were both based on two-photon detection\cite{zhao2007robust,chen2007fault}. In 2008, N. Sangouard et al. proposed a scheme to generate robust and efficient entanglement of two quantum network nodes based on two-photon detection and partial readout of ensemble memories with four CAEs\cite{sangouard2008robust}. To experimentally realize two-photon based entanglement, S. Chen et al. demonstrated robust entanglement between two spatial modes of collective excitations and two polarization states of photons in 2007, where the Fock-state qubit $|0\rangle_a$ was no longer used\cite{chen2007demonstration}. Based on this type atom-photon entanglement source, in 2008, Z. S. Yuan et al. reported the entanglement of two CAEs with two-photon detection (See Figure \ref{fig:6}), where the negative effect of state $|0\rangle_a$ is taken out of consideration\cite{yuan2008experimental}. 
In their experiment, two pairs of atom-photon entanglement were generated through a protocol described in Ref. \cite{chen2007demonstration}, the two photons were sent to a polarization-distinguishing Bell-state analyzer\cite{pan2012multiphoton}. With a specified coincidence detection of two photons, the spatial modes of the collective excitations in two CAEs would collapse to an entangled state. In the whole experiment, to avoid multi-excitations of atoms and emission of multi-photons, short and low intensity write pulses had to be used, which constrained the excitation probability to the level of 0.01. As a result, the overall detection efficiency of write photon was around $2.5 \times 10^{-3}$. And the entanglement generation efficiency was $3.1 \times 10^{-6}$. The intrinsic retrieval efficiency was 0.35, and the transmission and detection efficiency was about 0.4 leading to overall detection efficiency of read-out photons of around 0.15. Finally, the entanglement verification efficiency was $7 \times 10^{-8}$. Considering the whole experimental repetition rate of 10 kHz, the overall count rate of coincidences was 0.04 per minute.

To acquire a stronger collective enhancement and larger excitation probability in a small-size atomic cloud, in 2007, J. Simon et al. tried to place the cold cesium atomic ensembles inside a standing-wave Fabry-Perot cavity with a finesse of 93 and realized an intrinsic retrieval efficiency of 0.84\cite{simon2007interfacing}. However, with a Fabry-Perot cavity, the photons from forward and backward direction needed complex filtering operations which complicated the experimental device. As a solution, in 2012, X. H. Bao et al. prepared cold rubidium atoms inside a ring cavity instead of Fabry-Perot cavity\cite{bao2012efficient}, and reported a quantum memory with an intrinsic retrieval efficiency of 0.73. In contrast to a standing-wave cavity, the ring cavity can help to distinguish photons from the two-opposite directions, while, unfortunately, etalon and atomic vapor cell still had to be kept in the setup to filter noise photons, which induced inevitable loss of photons. Without these extra filters, the final photon detecting efficiency of a quantum memory could be further improved. Accounting for this, in 2019, B. Jing et al. used the cavity itself as a filter and freed all extra filters by setting the write-out, read-out photons resonant with the cavity, while the write, read beam were off resonant with the cavity\cite{jing2019entanglement}. After selecting energy levels with larger transition CG coefficients for quantum memory, they finally realized a quantum memory based CAEs with an intrinsic retrieval efficiency of 0.88 and an overall detected retrieval efficiency $\eta_r$ of 0.40. As the retrieval efficiency is larger, a higher excitation probability up to 0.02 was allowed, since the signal-to-noise ratio of atom-photon non-classical correlation was mainly determined by the ratio of retrieval efficiency to excitation probability\cite{jing2019entanglement}. With these improvements, the overall generation and verification efficiency of atom-photon entanglement were also improved.

\begin{figure}[t]
\centering
\includegraphics[width=15cm]{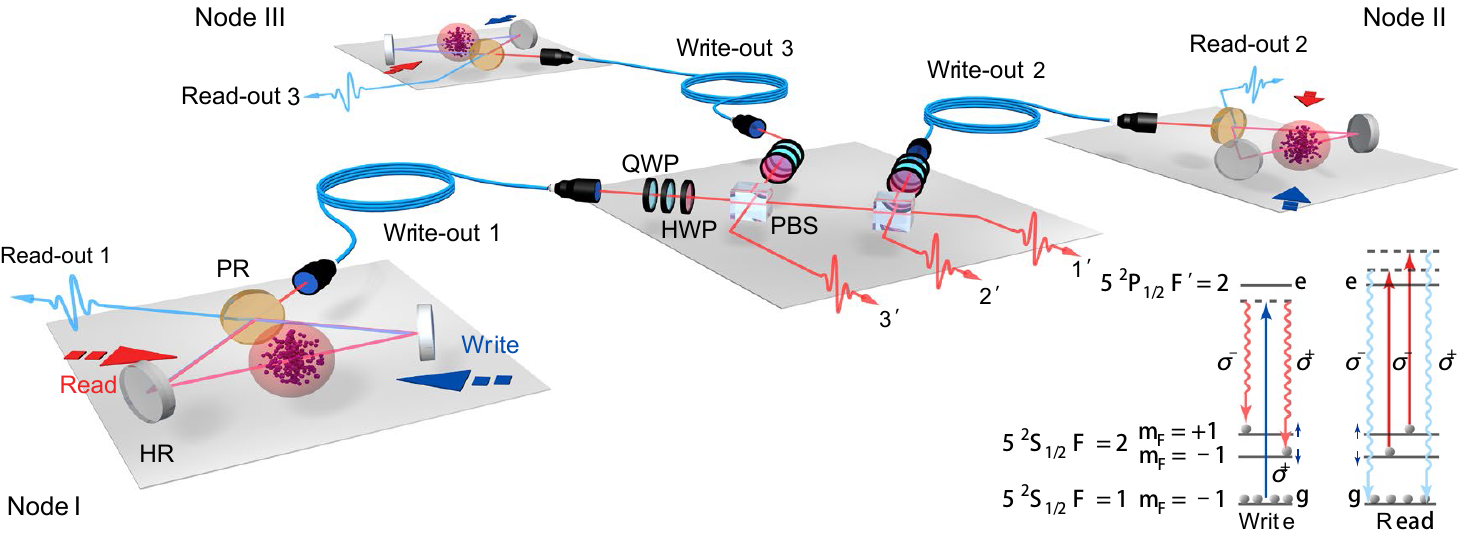}
\caption{Experiment setups for generating entanglement of three $^{87}Rb$ atomic ensembles. In each node, applying a $\sigma_{+}$-polarized write light pulse, the entanglement between Zeeman sublevel of atom and polarization of photon is created through spontaneous Raman scattering, where $\sigma_{-}$-polarized ($\sigma_{+}$-polarized) photon corresponds to the state $|F=2,m_{F}=1\rangle$ ($|F=2,m_{F}=-1\rangle$). The write out photons from three quantum nodes are sent to quantum interferometer to realize entanglement connection among three CAEs. the atomic collective states can be retrieved to optical states by applying a $\sigma_{-}$-polarized read light pulse and detected. Adapted from Ref. \cite{jing2019entanglement}.}
\label{fig:7}
\end{figure}

Similar to methods of atom-photon entanglement generation in single atoms, it is also possible to generate entanglement with Zeeman states instead of spatial modes in CAEs, which could further simplify the experimental devices. It was reported as early as in 2005, when D. N. Matsukevich et al. used atomic Zeeman states for encoding qubits and successfully prepared entanglement between photonic polarization states and orthogonal collective spin excitations\cite{matsukevich2005entanglement}. With this type of entanglement source, in 2006, they further transmitted and stored the photon in another CAEs, mapping the photonic states onto atomic states with EIT protocol\cite{matsukevich2006entanglement}, where a probabilistic generation of entanglement between two CAEs was reported. By implementing the entanglement generation process within a cavity to acquire a more efficient source, in 2015, S. J. Yang et al. reported this kind of atom-photon entanglement generation with high retrievable efficiency inside a cavity\cite{yang2015highly}. Utilizing similar experimental setup and filter-free operations, in 2019, B. Jing et al. reported generation of three separate pairs of atom-photon entanglement, and the three write-out photons were sent to a GHZ-state analyzer. With the detection of three photon coincidences, the three separate CAEs were entangled with a heralding efficiency of 14\%\cite{jing2019entanglement}, for details see Figure \ref{fig:7}. 
This three-fold GHZ state of three quantum memories represents the largest number of entangled quantum nodes within maximum entangled states up to now. While photon transmission in all experiments mentioned above was constrained to hundreds of meters, it was, nevertheless, an accomplishment to achieve this modest node distance. To reach much longer photon transmission distances, it is necessary to convert wavelength of the photons from the quantum memories into telecom-band photons for long distance. To reach much longer photon transmission distances, it is necessary to convert wavelength of the photons from the quantum memories to a wavelength that is optimized for the transmission channel. Usually, the wavelength of photons should be converted to the telecom band for the lowest loss over fiber channels. Actually, CAEs themselves can act as quantum frequency converters\cite{radnaev2010quantum,dudin2010entanglement}. Alternatively, difference-frequency generation processes in periodically poled lithium niobate waveguides can also be employed to convert the wavelength of photons to the telecom band\cite{yu2020entanglement}. In 2020, Y. Yu et al. combined this method with a high retrieval efficiency quantum memory and realized the entanglement of two cold rubidium atomic ensembles via photon transmission through city-scale optical fiber, i.e., 22 km fiber with two-photon detection and 50 km fiber with one photon detection. It is the longest distance of photon transmission to date over which entanglement of quantum memories has been generated. In 2021, Y. F. Pu et al. demonstrated entanglement between two $^{87}Rb$ atomic ensembles with the fidelity of 79.5\%. The lifetime of the quantum memories was tens of milliseconds, which has a positive effect on the scaling of the entanglement generation efficiency\cite{pu2021experimental}. It is worth noting, that free-space transmission, e.g., used in satellite-based quantum networks, is usually done with photons at wavelengths from 780-850 nm, which overlaps with the common emission wavelengths of rubidium (780 nm and 795 nm) and caesium (852 nm and 894 nm). Hence, for such applications no wavelength conversion would be required.
\subsubsection{Entangling quantum memories via other ways}
Besides of connecting quantum nodes based on DLCZ-type embedded quantum memories in CAEs, it is also possible to use absorptive memories such as EIT quantum memory, Raman quantum memory, Autler-Townes splitting (ATS) memory\cite{saglamyurek2018coherent}, etc., for which one member of a pair of externally generated entangled photons is stored. A storage efficiency > 85\% and a storage fidelity > 99\% with EIT have been reported\cite{wang2019efficient}. In 2015, D. S. Ding et al. realized Raman quantum memory of photonic polarized entanglement and entangled two CAEs\cite{ding2015raman}. In 2017, Z. H. Yan et al. stored multi-partite continuous-variable entanglement in three atomic ensembles and demonstrated entanglement of three CAEs\cite{yan2017establishing}. Compared with discrete variable entanglement, the fidelity of continuous-variable entanglement in this experiment was sensitive to photon loss and less robust against transmission loss in long distance quantum network.

As a short summary, for quantum networks consisting of DLCZ-type embedded quantum memories in CAEs, the approach is appealing, as the large number of atoms can not only provide strong interaction between atoms and photons but also improve the retrieval efficiency through collective enhancement. On the other hand, the disadvantages of the approach are that the atom-photon entanglement is generated probabilistically and the excitation probability has to be held low to suppress multi-excitations, which sets a bottleneck to improve the efficiency of the entanglement source. Fortunately, through Rydberg blockade, it is possible to generate deterministic atom-photon entanglement\cite{saffman2010quantum}, which will improve the excitation probability to unity and enable a much higher creation rate for the multi-nodes entanglement. It can also eliminate the multi-excitation events and improve the heralding efficiency after entanglement swapping operations. Moreover, the requirement for long memory lifetime during generation of long distance nodes entanglement has also been met\cite{yang2016efficient},in that, sub-second lifetime was demonstrated by confining atoms in an optical lattice. This would be a sufficiently long storage time to support hundreds of kilometers distance. As for the bandwidth, only a few MHz can be reached, as it is upper bounded by the collective linewidth of the optical transition involved in the coupling scheme. This relatively narrow bandwidth may make CAEs implementations less attractive as it limits the entanglement distribution rate.

\subsection{Quantum network with trapped ions}
Trapped ions can also act as quantum nodes for quantum networks\cite{duan2010colloquium} with unique set of advantages. First, single ions can be trapped easily for weeks and, thus, offer a much longer storage than that of single neutral atoms\cite{blatt2008entangled}. Second, the state of trapped ions can be detected near-deterministically by optical cycling transitions\cite{blatt1988quantum}. Third, the implementation of single-ion operations\cite{harty2014high,ballance2016high} and quantum memory\cite{wang2017single,sepiol2019probing} has high fidelity. Moreover, the theoretical schemes and methods applied to atomic systems can be also utilized in trapped ions systems. In 1998, Q. A. Turchette et al. demonstrated deterministic entanglement between two trapped ions in an elliptical radio-frequency Paul trap\cite{turchette1998deterministic}. Due to the probabilistic protocols that are feature greater robustness to noise, it is useful for establishing long distance quantum network. In 2003, C. Simon et al. described the potential of ions acting as matter qubits to generate ion-ion entanglement based on two-photon interference\cite{simon2003robust}. The initial states of both ions with $\Lambda$-type energy level structure are prepared to the excited state $|e\rangle$, and the ions will decay to two degenerate metastable states $|s_1\rangle$ and $|s_2\rangle$ through emitting photons with different polarizations. In this way, the maximal entanglement are generated between polarizations of emitted photons and internal states of an ion in each quantum node. Two photons emitted from two different ions are sent to a partial Bell-state analysis to perform entanglement swapping, leading to generation of entanglement between two ions. Furthermore, in 2009, N. Sangouard et al. analyzed the performance of this scheme, and developed a protocol to realize temporal multiplexing entanglement\cite{sangouard2009quantum}. According to the analyses in Ref.\cite{simon2003robust,sangouard2009quantum,luo2009protocols}, typically, there are two steps to implement heralded entanglement between two trapped ions. First, the entanglement between an ionic qubit and photonic qubit is implemented through the scattering of photons. Second, detecting results of the interference of the scattered photons from different ions projects the state of the two ions into an entangled state. 

\begin{figure}[t]
\centering
\includegraphics[width=13cm]{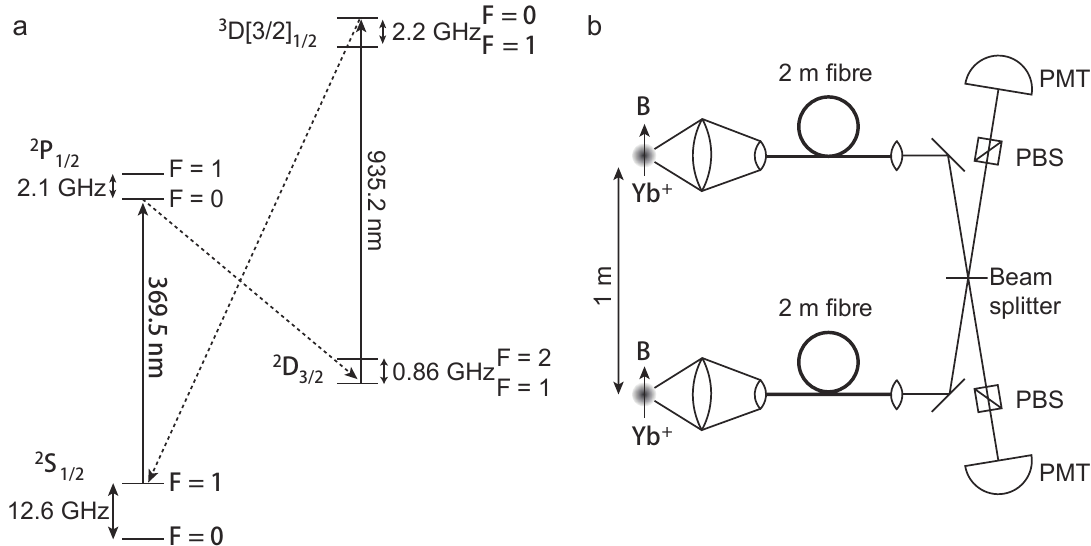}
\caption{Entanglement generation between two $^{171}Yb^{+}$. (a) The energy levels structure of $^{171}Yb^{+}$. The transition $^2$S$_{1/2}$ to $^2$P$_{1/2}$ is pumped by a light with the wavelength of 369.5 nm. While the ion is excited to $^2$P$_{1/2}$ level, it can decay to the $^2$D$_{3/2}$ level. And this level can be driven to the $^2$D[3/2]$_{1/2}$ level by a 935.2 nm light. (b) Experimental setups. The two trapped ions are placed into two independent vacuum chambers separated by 1 meter. And the imaging lenses are used to collect the spontaneously emitted photons from two ions. The applied magnetic filed B determines the polarization of each spontaneously emitted photon. Two photons from two ions are send to a BS and carried out Bell-state measurement, forming entanglement between two ions. Adapted from Ref. \cite{moehring2007entanglement}.} 
\label{fig:8}
\end{figure}

Generally, for any quantum network with matter qubits, one basic requirement is the generation of light-matter entanglement, and this is also true for ions. In 2004, B. B. Blinov et al. reported quantum entanglement between hyperfine ground states of a single trapped $^{111}Cd^{+}$ ion and polarization states of a single photon emitted spontaneously from the ion\cite{blinov2004observation}. As a proof-of-principle experiment, due to the single photon emission efficiency of 0.1 and the single photon detection efficiency of $1.6\times10^{-3}$, the entanglement generation efficiency was $1.6\times10^{-4}$. The experimental repetition rate was 2 kHz, leading to an entanglement generation rate of 18 per minute. At the same year, D. L. Moehring et al. performed a measurement of a Bell inequality violation with this entanglement and demonstrated violation of the Bell inequality between particles of different species\cite{moehring2004experimental}. After the preparation of single ion-single photon entanglement, the photon can be used as a quantum bus to connect different nodes containing ions. In this spirit, in 2007, D. L. Moehring et al. demonstrated the entanglement generation between two separated $^{171}Yb^{+}$\cite{moehring2007entanglement}. In their experiment, two ions were trapped in two vacuum chambers separated by 1 meter. The experimental apparatus is displayed in Figure \ref{fig:8}.
To entangle two separate ions, the first step is to produce entanglement between internal state of single ions and the spontaneously emitted photons. A $\sigma^{-}$-polarized ultrafast laser pulse resonant with $^{2}S_{1/2}$ to $^{2}P_{1/2}$ electronic transition excited the ion, which was prepared into the state $|F,m_{F}\rangle=|0,0\rangle$ to exciting state ${^2}P_{1/2}$ $|1,-1\rangle$. Here $F$ is the total angular momentum and $m_F$ is its projection along the quantization axis which is determined by the applied magnetic fields. Through the spontaneous emission of single photon with different polarizations, the state ${^2}P_{1/2}$ $|1,-1\rangle$ was transferred to different hyperfine levels of ground. State ${^2}S_{1/2}$ $|1,-1\rangle$ corresponds to a $\pi$-polarized photon and state ${^2}S_{1/2}$ $|1,0\rangle$=$|\uparrow\rangle$ and ${^2}S_{1/2}$ $|0,0\rangle$=$|\downarrow\rangle$⟩ correspond to a $\sigma^{-}$-polarized photon. By filtering the $\pi$-polarized photon, the entanglement between ion-photon was generated,
\begin{equation}
|\psi\rangle_{ion-photon}=\frac{1}{\sqrt{2}}(|\uparrow\nu_{\uparrow}\rangle-|\downarrow\nu_{\downarrow}\rangle),
\end{equation}
 where $|\nu_{\uparrow}\rangle$ and $|\nu_{\downarrow}\rangle$ represent the two resolved frequencies of ground-state hyperfine splitting ${^2}S_{1/2}$ $|1,0\rangle$ and ${^2}S_{1/2}$ $|0,0\rangle$. The second step is to send the two photons from different ions on a BS and carry out the Bell-state measurement. Through appropriate coincidence detection of photons, state of two ions would collapse to an entangled state, 
 \begin{equation}
|\Psi^{-}\rangle_{ion-ion}=\frac{1}{\sqrt{2}}(|\uparrow_1\downarrow_2\rangle-|\downarrow_1\uparrow_2\rangle).
\end{equation}
Finally, an entanglement fidelity of 63\% was acquired. As the first demonstration of separated ion-ion entanglement, the entanglement generation efficiency was $3.6\times10^{-9}$. With an experimental repetition rate of 550 kHz, the entanglement generation rate was about 0.12 per minute. 

As extensions of Ref. \cite{moehring2007entanglement}, there were several experiments carried out with two separated trapped $^{171}Yb^{+}$ ions and their non-classical correlated photons\cite{matsukevich2008bell,olmschenk2009quantum,maunz2009heralded,vittorini2014entanglement}. In 2008, D. N. Matsukevich et al. demonstrated the entanglement between two $^{171}Yb^{+}$ ions separated by 1 meter and further verified Bell inequality violation with ion-ion entanglement\cite{matsukevich2008bell}. The experiment in Ref. \cite{matsukevich2008bell} differed from that in Ref. \cite{moehring2007entanglement}, in which the qubit was encoded in the frequency of the emitted photon, by, instead, utilizing the polarization degree of freedom for the qubit encoding. Two pairs of ion-photon polarization entanglement were generated. After photons passed through a Bell-state analyzer, with coincidence detection, the entanglement between two ions were generated with a fidelity of 81\% and an entanglement generation rate of 1.56 per minute. The entanglement fidelity was improved compared to Ref. \cite{moehring2007entanglement}. This is due to utilization of the more robust polarization degree of photons. The measurements of both ion and photon can be carried out in arbitrary bases and the ion-photon and ion-ion entanglement can be characterized by quantum state tomography. The entanglement generation rate was improved 13-fold through a different excitation scheme that capable of transferring all population to the excited state and through a different photon collection scheme that does not require polarization filters. In 2009, S. Olmschenk et al. teleported a qubit stored in a single trapped ytterbium ion to a second ytterbium ion with an average fidelity of 90\%\cite{olmschenk2009quantum} using a teleportation protocol based on heralded entanglement between the two ions. In their experiment, ion-photon frequency entanglement was generated and entanglement swapping was operated through interference of photons. Finally, an ion-ion entanglement generation efficiency of $2.2\times10^{-8}$ was accomplished. Due to the experimental repetition of 75 kHz, the successful teleportation rate was about 0.083 per minute. At the same year, P. Maunz et al. performed a heralded quantum gate between remote quantum memories consisting of single $^{171}Yb^{+}$ ions and entangled them together\cite{maunz2009heralded}. The maximum efficiency of entangling gate generation was about $4.2\times10^{-8}$. The experimental repetition was 70 kHz, leading to the entanglement generation rate of 0.09 per minute. Combined with time-resolved photon detection, in 2014, G. Vittorini et al. experimentally generated heralded entanglement between two remotely trapped $^{171}Yb^{+}$ ions via interference of two distinguishable photons\cite{vittorini2014entanglement}, which presented a feasible way for future quantum network that are composed of heterogeneous qubits.

From the progresses shown in the above works, the generation efficiency of ion-ion entanglement is still too low to achieve a scalable quantum network using trapped ions, and one of the key factors is the photon collection efficiency. Usually, this issue can be solved by placing the trapped ions inside a cavity as done in single atom experiment. In 2012, A. Stute et al. demonstrated tunable entanglement between a single $Ca^+$ ion and polarization state of a single photon in an optical cavity\cite{stute2012tunable}, for which the probability of detecting a photon in a single sequence was 5.7\% and the ionic state could be measured with a 100\% efficiency, thus a final ion-photon entanglement with an overall generation and verification efficiency of 5.7\% was reported. It represented the highest efficiency among all light-matter entanglement up to then and provided an efficient source supporting multi-nodes entanglement. Similar to the single neutral atom setups, it is possible to avoid using a cavity, by instead employing collection lens with a much larger N.A. (0.6), as was done by D. Hucul et al. in 2015\cite{hucul2015modular}. In their experiment, modular entanglement of atomic qubits using photons and phonons was created. Two $Yb^+$ ions located in separated modules were connected through ion-photon entanglement and photon interference with efficiency of $9.7\times10^{-6}$. Heralded entanglement with a fidelity of 78\% and a generation rate of 4.5 per second was achieved. While for ions within the same module, the Coulomb-coupled transverse phonons modes of the atoms and deterministic near-field interaction through phonons gave a possible way to create entanglement. The two kinds of ion-ion entanglement generation presented in their demonstration suggested an experimental route for scalable quantum network, which consists of multiple modules with each module using several interconnection-available single ions as quantum memories. In 2020, L. J. Stephenson et al. reported a novel excitation scheme using $^{88}Sr^+$ ions with photon collection perpendicular to the applied static magnetic field, resulting in an increased entanglement generation rate over previous experiments\cite{stephenson2020high}. Through the use of a high N.A. (0.6) lens, a final heralded ion-ion entanglement with fidelity of 94\% was generated. The successful efficiency of entanglement was about $2.18\times10^{-4}$, and due to the experimental repetition of 833 kHz, the heralded photon entanglement generation rate was about 183 per second. Besides generating ion-ion entanglement with two-photon detection, which is more robust against phase fluctuations, the two separate ions can also be encoded with Fock-state qubits and entangled via single-photon detection, which can lead to an even higher entanglement generation rate using the same ion-photon source. Such an experiment was carried out by L. Slodi\v{c}ka et al.\cite{slodivcka2013atom} and the probability of preparing an entangled state with single-photon detection was $1.1\times10^{-4}$, leading to a final entanglement generation rate of 14 per minute.

For a brief summary, trapped ions can be easily confined and addressed for much loner time, and significantly longer operation times for ion qubits are allowed. The measurement of ionic states can also reach near-unity efficiency, which shows great potential for the detection process of entanglement. Furthermore, the unique advantages of ions as quantum network nodes are the local convenience and deterministic Coulomb-based gates between adjacent ions, where no photon conversion is needed. Using such gates, 14-qubits entanglement\cite{monz201114} and 20-qubits entanglement\cite{friis2018observation} have been experimentally observed with a string of $Ca^{+}$ ions confined in a linear Paul trap. However, for the entanglement generation between physically separated ions memory nodes, photons are still needed as information carriers, and thus ion-photon entanglement is the basic requirement. A cavity-based entanglement generation process or the use of large N.A. collection lenses provides an efficient way to prepare a high-brightness entanglement source, which is of great significance for entangling distant quantum nodes.

\subsection{Quantum network with Nitrogen-Vacancy center in diamonds}
\begin{figure}[htp]
\centering  
\includegraphics[height=10.5cm]{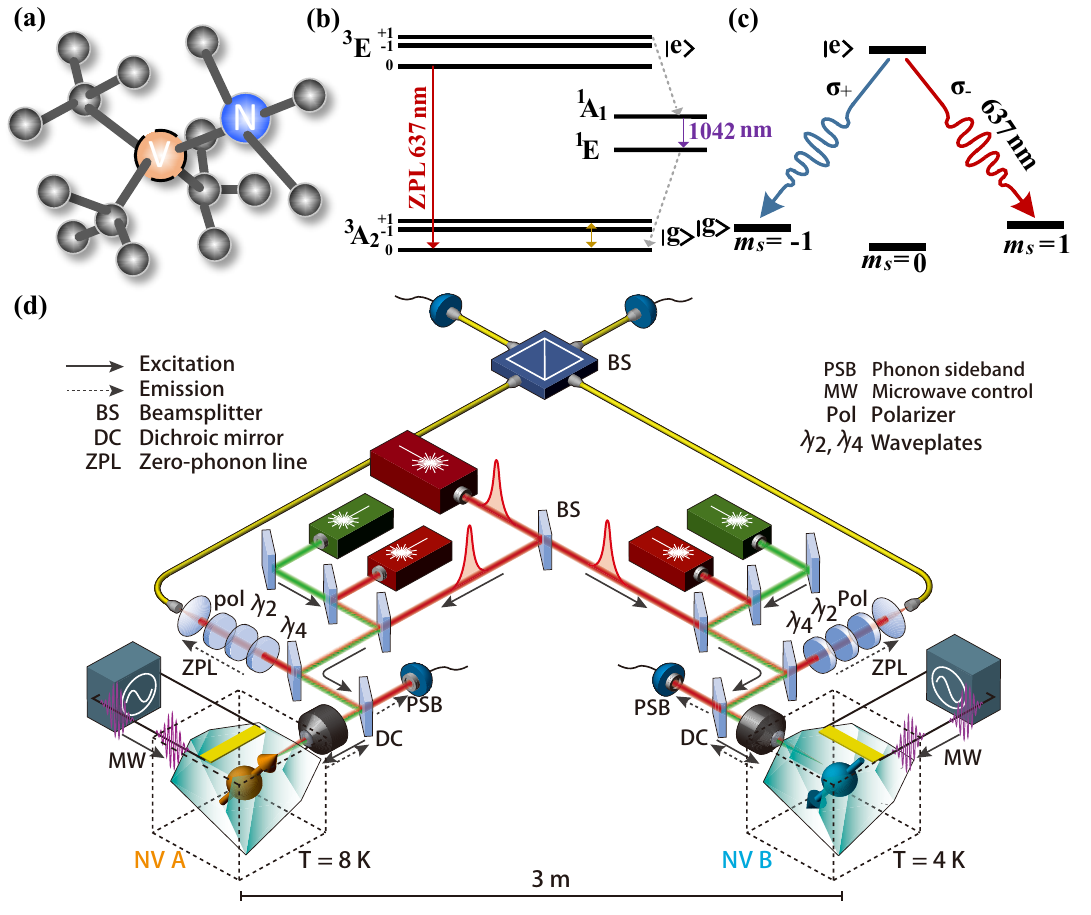}
\caption{The NV center in diamond. (a) Schematic of an NV center in diamond. The NV electron spin is coupled to proximal nuclear spins, such as its intrinsic N and lattice C. (b) The basic energy level structure. The crystal field splits the triplet sub-levels of the ground and excited state. The zero-phonon line (ZPL) transition is indicated. The dashed arrows indicate the dominant transitions to and from the metastable shelving states which enable the optical spin polarisation mechanism. (c) The scheme of energy level for generation of entanglement. NV center is excited to state $|A_2\rangle$, after spontaneous emission of single photon with possibly different polarizations, the excited state is transferred to a long-lived spin state ($|m_{s}=\pm1\rangle$). The spin state $|m_{s}=1\rangle$ corresponds to a $\sigma_-$-polarized photon and $|m_{s}=-1\rangle$ corresponds to a $\sigma_+$-polarized photon. (d)The entanglement generation between two NV centers. Two diamonds are placed into two independent microscope apparatus separated by 3 m. The red laser can resonantly excite the NV centers, and the green laser can excite them off-resonantly. The emission represented by dashed arrows is separated into phonon sideband (PSB) and zero-phonon line (ZPL). The PSB is utilized for spin qubits independent readout, and the ZPL photons of two NV centers are combined on a beam splitter (BS). The spin is controlled by applying microwave pulses. The m$_{S}$=$\mp1$ energy levels are split by applying a 17.5 G magnetic field, and a DC electric field is used to tune the optical frequencies of NV B. Firstly, the NV centers are prepared in the sate of $\frac{1}{\sqrt{2}}(\lvert\uparrow\rangle+|\downarrow\rangle)$. Then the spin-photon entanglement state of $\frac{1}{\sqrt{2}}(\lvert\uparrow1\rangle+\lvert\downarrow0\rangle)$ is created by a 2 ns long resonant laser pulse. The photons are combined on the BS and detected after the BS. The entanglement generation is heralded by detecting one photon. Figure (d) adapted from Ref. \cite{bernien2013heralded}.}
\label{fig:9}
\end{figure}

Cold atomic gasses, single atoms, or ions have all proven to be promising systems for encoding qubits and serving as quantum memories. However, quantum devices based on these platforms would appear very different from those used in current information technologies. Solid-state platforms, however, resemble current information technology much closer, and could inherit the benefits of scalable, integrated, and const-effective fabrication\cite{gao2015coherent}. For instance, Nitrogen-Vacancy (NV) color centers in diamond have been widely used in quantum information processing\cite{awschalom2018quantum}. The NV center is a point defect in the diamond lattice, and it consists of a nearest-neighbor pair of a nitrogen atom, which substitutes for a carbon atom and a lattice vacancy, as shown in Figure \ref{fig:9}(a). The energy level structure of NV center is shown in Figure \ref{fig:9}(b). The ground state of NV center is a spin-triplet ground state $(|m_{s}=0\rangle,|m_{s}=-1\rangle,|m_{s}=1\rangle)$), which can provide long spin coherence lifetime. The zero-phonon line (ZPL) of the transition between spin-triplet ground state and excited state is 637 nm, and the microwave frequency corresponding to zero field splitting between spin states $|m_{s}=0\rangle$ and $|m_{s}=\pm1\rangle$⟩ is 2.88 GHz. Therefore, the spin-triplet ground state could be utilized to encode qubits in quantum information science.

What attracts people mostly to use NV centers as quantum nodes is its available control at room temperature, long coherence time of the nuclear spin and controllable quantum gates between nuclear and electronic spin qubits\cite{childress2006coherent}. These features makes NV-centers very suitable for not only quantum computing but also quantum networks. Here we mainly consider the latter. As reported in 2007 by M. V. Gurudev Dutt et al., the nuclear spin of an NV center could be used as a room-temperature quantum memory to store prepared electronic spin qubits for several microseconds\cite{dutt2007quantum}. They demonstrated an efficient way for operation on nuclear and electronic spin qubits. Although it is possible to maintain the spins of NV centers at room temperature, the decoherence and ZPL are, nevertheless, adversely affected by temperature. Thus, a low temperature environment both improves the coherence time and inhibits phonon relaxation process efficiently. For this reason, most experiments on NV centers are carried out at the temperature ranging from a few milli-Kelvin to several Kelvin.

For all the quantum networks based on entanglement of different nodes, it is necessary to implement a light-matter interface such that photons can be employed as flying qubits to connect distant matter nodes. In 2004, L. Childress et al. presented a fault-tolerant protocol for constructing a quantum repeater network in a solid-state environment\cite{childress2006fault}. In this scheme, the nodes consist of single photon emitters in solid-state systems, such as NV center in diamond and quantum dots, where the nuclear spin are used to store quantum information, and electronic spin is utilized for communication with adjacent nodes. They took NV center as an example to further describe the possible realization of entanglement connection using these solid-state materials. In NV center, the entanglement connection is realized in electronic triplet ground and the nuclear spin state of a nearby $^{13}C$. Therefore, single NV center with a nearby $^{13}C$ can act as a quantum node.  According to spontaneous decay to a certain electronic spin state (such as ground state $|m_{s}=-1\rangle$ or $|m_{s}=1\rangle$), the NV center can emit photons with different polarizations. Two photons from adjacent NV centers are transmitted to a interferometer and Bell-state measurement is performed. According to the measured results, the nuclear spin states of two NV centers can be projected onto an entangled state, i.e., entangled quantum nodes can be realized via this way. Later in 2005, they further presented the details of the protocol\cite{childress2005fault} which has the lowest requirements on physical resources for fault-tolerant quantum repeater. Several years after the proposals, the experimental generation of entanglement between photons and NV centers in diamond was experimentally realized by E. Togan et al. in 2010\cite{togan2010quantum}. The relevant energy level structure is shown in Figure \ref{fig:9}(c). The NV center was prepared in excited state $|e\rangle$, and through spontaneous emission of a single photon with different polarization states the excited state is transferred to a long-lived spin state of ground state ($|m_{s}=\pm1\rangle$). The spin state ($|m_{s}=1\rangle)$) corresponds to $\sigma_-$-polarized photon and ($|m_{s}=-1\rangle$) corresponds to a $\sigma_+$-polarized photon. In this way, entanglement between the NV center and the single photon is generated,
\begin{equation}
|\Psi\rangle_{AB}=\frac{1}{\sqrt{2}}(|\sigma_-\rangle|m_{s}=1\rangle+|\sigma_+\rangle|m_{s}=-1\rangle).
\end{equation}
Another useful property of the NV spin states is that certain excited states only couple to the spin state $|m_{s}=0\rangle$, which thus constitutes a cycling transition. Hence, it is possible to measure the $|m_{s}=0\rangle$ population through fluorescence detection To measure the qubit state encoded in the spin state $|m_{s}=\pm1\rangle$, a $\pi$ pulse can be used to individually transfer the states $|m_{s}=1\rangle$ or $|m_{s}=-1\rangle$ to $|m_{s}=0\rangle$ and detect the scattered photons through on the cycling transition. This scheme allows near-unity measurement efficiency of the qubit state. As a next step, in 2013, H. Bernien et al. demonstrated heralded entanglement between two NV center spins separated by 3 meter\cite{bernien2013heralded}. This experiment started with establishing entanglement between the spin state of each NV center and the photon number state of spontaneous emission. Then the indistinguishable spontaneously emitted photons from each NV center were combined by a BS erasing the path information and projecting the NV centers into an Fock-basis entangled state by detection of a single photon, see Figure \ref{fig:9}(d).
The probability of generating entanglement based on the single-photon detection scheme is usually low, and in this experiment, it was about $10^{-7}$, which is determined by overall detection efficiency of resonant photons from two NV centers. With an experimental repetition of 20 kHz, the entanglement generation rate was about 0.1 per minute. And the fidelity of entanglement was 92\%.

As an extension and application of this work, in 2014, W. Pfaff et al. performed a quantum gate between the nuclear and electronic spin state and demonstrated quantum teleportation between two NV centers separated by 3 meter\cite{pfaff2014unconditional}. This experiment started with generating entanglement between two remote electron spins of NV centers (NV A and NV B) using the method of Ref. \cite{bernien2013heralded}. The state to be teleported was provided by nitrogen-14 (N-14) nuclear spin in the vicinity of NV A. Hence, a Bell-state measurement between the N-14 nuclear spin state and the electron spin state of NV A was carried out and the measurement result was sent to NV B. Finally, NV B performed correcting transformation according to the measurement results to obtain the state to be teleported. The fidelity of teleported state was about 86\%. As mentioned earlier, the electron spin state can be measured with a near-unity efficiency, which can close the detection loophole when verifying Bell inequality. Therefore, in 2015, B. Hensen et al., used this setup to verify the loophole-free Bell inequality violation via two entangled NV centers separated by 1.3 kilometers\cite{hensen2015loophole}. This seminal affirmation of non-locality is at the same time the longest distance of physical separation of entangled quantum nodes reported so far. A great advantage for NV centers is that efficient quantum gates can be operated directly between nuclear and electronic states. For example, entanglement can first be established between two separate NV centers via photon interference of emitted photons. Then this entanglement can be transferred to the nuclear spin states in of carbon-13 (C-13) atoms, which are part of the diamond lattice close to the NV centers. After erasing the qubit information of the NV centers electronic states, they can be utilized to generate entanglement again, i.e. two pairs of entangled states can be prepared with either electronic spin states of the NV or nuclear spin states of C-13. By using the two pairs of entanglement located in two nodes, entanglement purification can be applied to acquire a higher fidelity of entanglement between two NV centers with the sacrifice of one entanglement pair. In 2017, N. Kalb et al. demonstrated this entanglement distillation process between two NV centers[\cite{kalb2017entanglement}. This approach also greatly reduces the requirement of phase-stability of the entire channel when using the Fock basis to encode the qubits. This is due to the cancellation of the global phase, when performing the entanglement distillation using the two pairs of entangled spins. In 2018, P. C. Humphreys et al. demonstrated deterministic delivery of remote entanglement on a quantum network\cite{humphreys2018deterministic}, in which single-photon detection was used, and a node-node entanglement generation rate of up to 39 Hz was reported. This , which was three orders of magnitude higher than two-photon protocols previously demonstrated\cite{pfaff2014unconditional} on the same platform. Recently, M. Pompili et al. reported realization of entanglement among three NV centers and any-to-any connection through entanglement swapping\cite{pompili2021realization}.

As a brief summary, quantum gates between nuclear and electronic spin states in NV centers have been experimentally demonstrated and have paved a path towards entanglement swapping with local operations and entanglement generation with long coherence time. To accounting for the longer operation times, encoding quantum states into a decoherence-protected subspace of two nuclear spins has also reported. In this fashion quantum coherence could be maintained for over 1000 repetitions of the remote entanglement protocol\cite{reiserer2016robust}. Even so, with currently reported experimental parameters, it is difficult to entangle more distant quantum nodes. As in the other physical systems mentioned above, the wavelength of photons correlated with matter qubits is in the infrared or near-infrared band, leading to higher transmission loss compared to telecom-band photons. Development and use of efficient quantum frequency conversion remains essential in order to extend the physical separation of quantum nodes\cite{zaske2012visible,dreau2018quantum}. Furthermore, as a unique weakness in NV-like solid state materials, the narrow ZPL significantly affects the photon emission probability, as most qubit information in the NV centers relax via combined phonon and spontaneous emission of photons outside the ZPL, which are not detected and in any case not correlated with the NV spin-state. As a result, photons from NV centers feature an extremely low efficiency for final detection. It is challenging to inhibit the phonon relaxation process and improve the photon emission efficiency via the ZPL. One approach is to couple the NV centers to cavities so as to enhance the ZPL via the Purcell effect\cite{bogdanovic2017robust,jung2019spin}.

Another tactic is to investigate and use new types of vacancy defects in diamonds, which may provide better optical properties. Recently, SiV center and SnV center have attract attention and are under thorough investigation\cite{nguyen2019integrated,nguyen2019quantum,bhaskar2020experimental,iwasaki2017tin,rugar2019characterization,gorlitz2020spectroscopic}. Defect centers in Silicon have also been found to emit through narrow zero-phonon lines at telecom wavelengths and could employ well-established silicon-photonics fabrication processes\cite{bergeron2020silicon,redjem2020single}.

\section{Quantum network with absorptive quantum memories}
The trend in the platforms discussed in preceding sections was that node-to-node entanglement generation employed embedded quantum memory from which emitted photons are entangled with an internal state of the memory. An alternative to this paradigm is instead to store externally provided, e.g., by a spontaneous parametric down-conversion process, entangled photons in two separate absorptive-type quantum memories and thus entangle these memories. Absorptive quantum memories can be realized on a number of platforms, but are typically ensemble based. One such ensemble-based solid-state memory is implemented in rare-earth ion doped solids (REIDS). Memory protocols in REIDS make use of the inhomogenously broadened transition lines to realize a form of photon-echo process, using external fields, spectral tailoring or a combination these. The REIDS platform is a favorable candidate for quantum nodes due to the long optical and spin coherence lifetimes by the shielded 4f-4f transitions of rare-earth ions\cite{bottger2009effects,zhong2015optically,ranvcic2018coherence}. However, even with the current state of the art, the efficiency of successfully preparing light-matter entanglement with REIDS is relatively low as compared to that with atoms or NV centers. Nevertheless, the greatest potential of REIDS acting as quantum memories is its ability to allow broadband and multi-mode quantum memory\cite{saglamyurek2016multiplexed,kutluer2017solid,yang2018multiplexed,seri2019quantum}, which provides an invaluable path for high-rate entanglement distribution. 

In the past two decades REIDSs have been widely studied for quantum information science and in particular for quantum memory. Excellent progresses has been made, such as a storage time of greater than one second\cite{longdell2005stopped}, high storage efficiency of 56\%\cite{sabooni2013efficient}, high storage fidelity of 0.999\cite{zhou2012realization} and simultaneous storage of up to 1060 temporal modes\cite{bonarota2011highly}. In addition, REIDSs have been shown to function as on-chip waveguide based quantum memory\cite{corrielli2016integrated,zhong2017nanophotonic,seri2018laser,liu2020reliable,zhu2020coherent,craiciu2021multifunctional,kindem2020control}, which is favorable for the construction of integrated quantum network. In the following we will focus on the atomic frequency comb (AFC) protocol\cite{afzelius2009multimode,de2008solid}, which distinguishes itself by it’s high-multimode storage capacity and use in most practical demonstrations using REIDS for quantum network applications.

\begin{figure}[htbp]
\centering
\includegraphics[width=11cm]{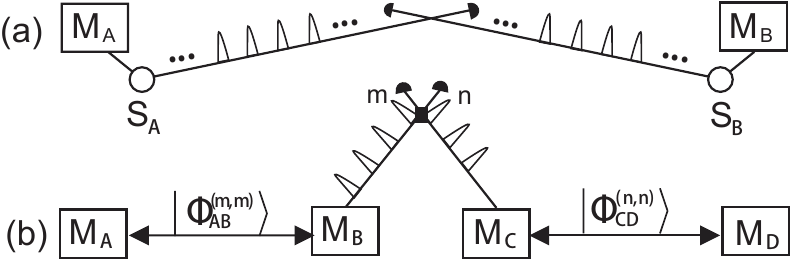}
\caption{Long-distance quantum networks based on entangled sources and multi-mode quantum memories. (a) The entangled photon-pairs sources emit a photon pair into a sequence of time bins at each node. One is stored into memory $M_A$ ($M_B$), and another is transferred to a BS at the middle of the link. Through detecting a single photon after the BS, the memories would be projected into an entanglement state. (b) Long distance quantum networks formed by entanglement swapping. If entanglement between the $m$th time bins in $M_A$ and $M_B$ and between the $n$th time bins in $M_C$ and $M_D$ are generated, by recovering photonic modes from memories and interfering on a BS, the entanglement between the $m$th time bins in $M_A$ and the $m$th time bins in $M_D$ are created. Adapted from Ref. \cite{simon2007quantum}.}
\label{fig:10}
\end{figure}

There are many protocols proposed to entangle remote quantum memories with REIDS. One typical scenario is about the realization of quantum repeater with photon pair sources and multi-mode memories proposed by C. Simon et al. in 2007\cite{simon2007quantum}, in which REIDS were employed as the quantum memories. In this protocol, one source of photon pairs and one memory are required at each node. The source is excited with a small probability to create a pair of photons, and one of the photons is mapped into quantum memory and another is transferred through a communication link. The photons from two quantum nodes A and B are then connected by the communication links, to beam splitter (BS) is inserted halfway between the nodes. Hence, the photons from node A and B are enter each one of the input ports of the BS and when a single photon is detected at one of the output ports of the BS, it is impossible to tell from which of the two nodes it originated. As a result, this detection heralds the state of the two nodes to become an entangled superposition in the Fock-basis,
\begin{equation}
|\psi\rangle_{AB}=\frac{1}{\sqrt{2}}(|1\rangle_{A}|0\rangle_{B}+|1\rangle_{A}|0\rangle_{B}),
\end{equation}
 where $|1\rangle_{A(B)}$ represents the state storing a single photon and $|0\rangle_{A(B)}$ represents the vacuum state. Thus, the elementary links from node A to node B would be formed, as shown in Figure \ref{fig:10}(a). To extend the distance of entanglement distribution, entanglement swapping should be applied on the basis of two nodes entanglement as described in DLCZ protocol. The setup of entanglement swapping is shown in Figure \ref{fig:10}(b). To create entanglement between node A and D, the memory excitation of node B and C are converted back into propagating photonic modes and transferred to communication links. Then these two communication links connect with the two input ports of a BS. When an photon is detected at one of the output ports of the BS, the entanglement swapping is successful. Finally, long-distance quantum nodes A and D are entangled. 
 
 As a beginning of the protocol, the storage of one of entangled photons forming light-matter entanglement is the first step. In 2011, C. Clausen et al. demonstrated the storage of energy-time entangled photons and confirmed that the non-classical properties are not changed after storage in $Nd^{3+}$:$Y_{2}SiO_{5}$\cite{clausen2011quantum}. This was verified by showing that the photon-pair, after storage of one of the members, still violated the Clauser-Horne-Shimony-Holt Bell-inequality\cite{clauser1969proposed} by three standard deviations ($S=2.64\pm0.23$). In a concurrent experiment, E. Saglamyurek et al. demonstrated storage of time-bin entangled photons and created photon-atom entanglement with a fidelity of 95\% in $Ti^{3+}:Tm^{3+}:LiNbO_{3}$ waveguide\cite{saglamyurek2011broadband}. In 2015, E. Saglamyurek et al. further realized storage of telecom C-band time-bin entangled photons and create photon-atom entanglement in $Er^{3+}$-doped silica-fiber\cite{saglamyurek2015quantum}. This experiment marked the entanglement was created between C-band light and matter. In 2016, A. Tiranov et al. demonstrated multi-temporal modes storage of entangled photons in $Nd^{3+}$:$Y_{2}SiO_{5}$ crystal\cite{tiranov2016temporal}. These demonstrations are all of great significance for realizing future quantum network.

To realize long-distance quantum network with absorption-type quantum memories, entanglement of two quantum memories is necessary. In 2012, I. Usmani et al. demonstrated entanglement between two neodymium ensembles doped in two yttrium orthosilicate crystals ($Nd^{3+}$:$Y_{2}Si0_{5}$) separated by 1.3 cm\cite{usmani2012heralded}. In their experiment, through spontaneous parametric down-conversion (SPDC), entangled photon pairs were generated, where the wavelength of idler/signal photon was 1338/883 nm. Then the idler photon was utilized as a heralding signal and the signal photon was sent to a BS, forming a single-photon entangled state between the two spatial out-put modes of the BS. Finally, the quantum state was stored into $Nd^{3+}$:$Y_{2}Si0_{5}$, and entanglement between two $Nd^{3+}$:$Y_{2}Si0_{5}$ crystals is generated. In this experiment to entangle two REIDS, a Fock-state qubit and single-photon detection were used. However, the whole entanglement generation and verification efficiency per experimental trial ($\sim$ $6.6\times10^{-5}$) wass still too low to extend distant node-to-node entanglement. 

\begin{figure}[htbp]
\centering
\includegraphics[width=16cm]{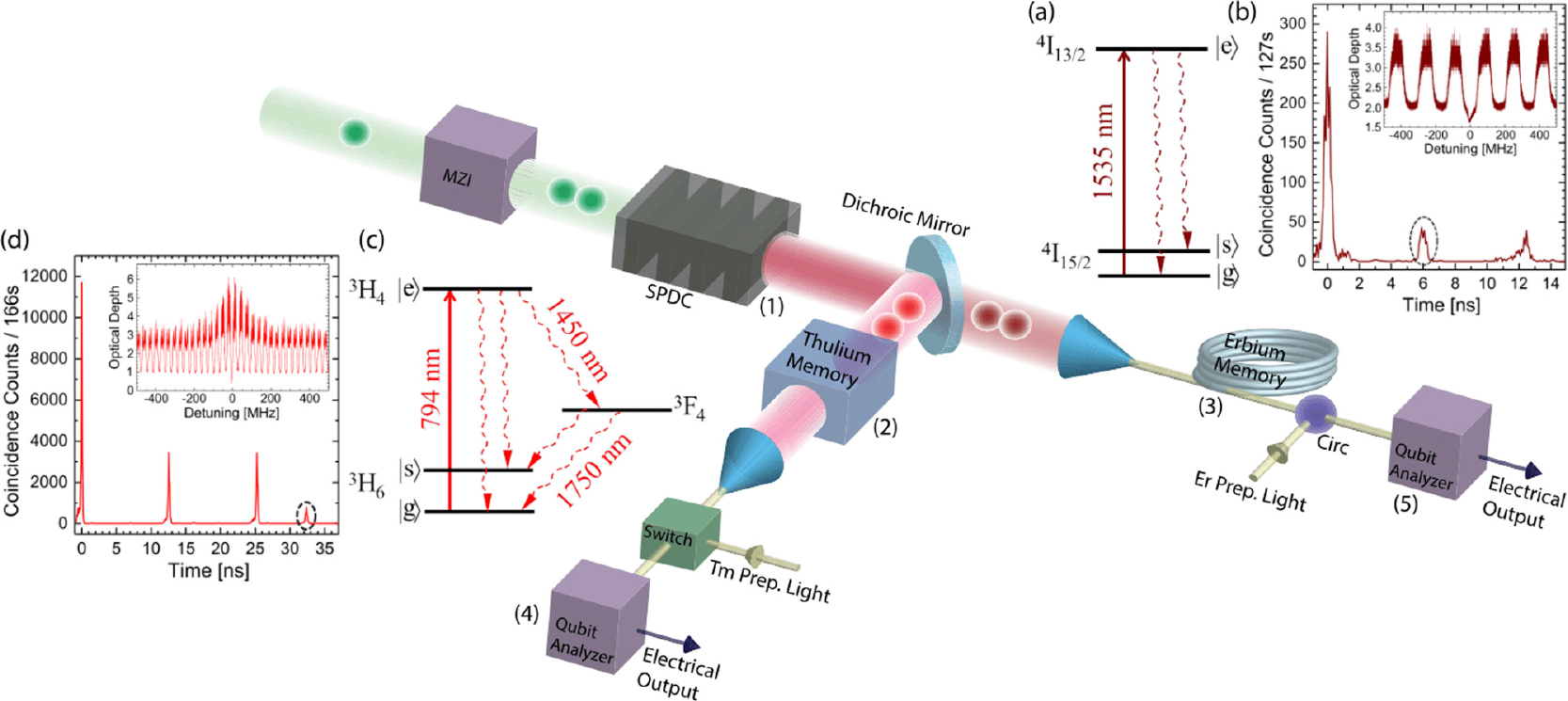}
\caption{Entanglement generation between two REIDSs. Time-bin entangled photon-pairs are transferred to two different REIDSs (erbium doped fibers and $Tm^{3+}$:$LiNbO_3$) and stored in REIDs forming entanglement between these two REIDSs and then re-emitted to verify this entanglement. (a) The erbium energy level structure. (b) The AFC prepared in erbium doped fibers with spacing of 166 MHz and storage time of 6 ns. The inset depicts an AFC based on erbium doped fibers with a section of 1 GHz wide. (c) The thulium energy level structure. (d) The AFC prepared in $Tm^{3+}$:$LiNbO_3$ with spacing of 31 MHz and storage time of 32 ns. The inset shows a AFC based on $Tm^{3+}$:$LiNbO_3$ with a section of 1 GHz wide. Adapted from Ref. \cite{Puigibert2019a}.}
\label{fig:11}
\end{figure}

With several years of development on quantum memories with REIDS, In 2020, M. L. G. Puigibert et al. experimentally demonstrated entanglement between two different solid-state quantum memories\cite{Puigibert2019a}, as shown in Figure \ref{fig:11}.
In their experiment, In their experiment, time-bin entangled photon-pairs were used, with wavelengths of the photons in each pair being 794 nm and 1535 nm. Then entangled photon-pairs of different wavelengths were separated by dichroic mirror. The photons with a wavelength of 1535 nm were stored into erbium doped fibers while 794 nm photons were stored into $Tm^{3+}$:$LiNbO_3$. Finally, the entanglement between two different REIDS was generated with input-output fidelity of 93\%. Due to the low storage efficiencies of 0.1\% in erbium doped fibers and 0.4\% in $Tm^{3+}$:$LiNbO_3$, the entanglement efficiency of two REIDS was about $4\times10^{-6}$. Although the entanglement efficiency was extremely low, this experiment has laid the ground to develop fiber-based future quantum network. Due to the low transmission loss of telecom photons in fiber, we can directly extend the distance of the entanglement by utilizing two telecom-memories, such as erbium doped fibers, $Er^{3+}:Y_{2}SiO_{5}$, $Er^{3+}:LiNbO_{3}$, etc.
The most recent demonstrations with REIDS have been two concurrent implementations of an elementary quantum repeater node. In one experiment by D. Lago-Rivera et al\cite{lago2021telecom}. Fock-basis entanglement was distributed by entanglement swapping at an impressive rate of 1.4 kHz into two $Pr^{3+}:Y_{2}SiO_{5}$ crystals separated by 10 m in different labs. The AFC storage time was set to 25 $\mu$s and heralded entangled photons were detected after storage at a rate of about 6 Hz. Importantly the heralding photons partaking in the Bell-state measurement were at telecom wavelength meaning that the link distance could be extended without much additional loss. However, the Fock basis entanglement requires phase stabilization of the entire link, which is a considerable challenge. Hence, only a simulated long-distance transmission was done using attenuators on the telecom wavelength channel before the Bell-state measurement. In the other experiment by X. Liu et al\cite{liu2021experimental} one member from two separate polarization entangled photon-pairs sources was interfered on beam splitter to perform a Bell-state measurement and thus swap the polarization entanglement to the two remaining photons, which were stored in separate $Nd^{3+}:YVO_{4}$ memories using the AFC protocol and spaced by 3.5 m. At a rate of 100 Hz the Bell-state measurement heralded entanglement in the memories featuring a storage time of 56 ns and 1 GHz bandwidth. Finally, based on the four-fold coincidence rate, the entanglement generation rate was $7.5 \times 10^{-12}$. Considering of the repetition of 40 MHz of photon storage, the entanglement distribution rate of two quantum memories is 0.3 mHz. The fidelity of the measured final entangled state was ($80.4\pm2.1$)\%.

In summary, for REIDS used for quantum networks, particularly, entanglement of separate quantum nodes, the modest performance achieved to date does not reduce this platforms significance. The most important reason is its unique capacity for broadband and multi-mode quantum memory. In recent years, great progress has been achieved in storage of photons with high fidelity\cite{zhou2012realization}, large bandwidth\cite{saglamyurek2016multiplexed} and multiple temporal modes\cite{tiranov2016temporal,usmani2010mapping,tang2015storage}, spectral modes\cite{saglamyurek2016multiplexed,sinclair2014spectral}, orbital-angular-momentum modes\cite{yang2018multiplexed,zhou2015quantum}, etc. As a solution for the low storage efficiency, a protocol employing an impedance-matched cavity has also been used to increase the storage efficiency\cite{sabooni2013efficient}, with a maximum efficiency of 56\% reported in 2013 by M. Sabooni et al. using $Pr^{3+}:Y_{2}SiO_{5}$. With regards to the storage time, which is initially determined by the prepared AFC, there are also methods to transfer the excitation to a long-lived spin transition, by which longer storage time and on-demand read-out can be realized. For the current state of the art, the longest storage time is 40 ms based on AFC protocol, which was realized by A. Holz\"{a}pfel with $Eu^{3+}:Y_{2}SiO_{5}$ in 2019\cite{holzapfel2019optical}. Another importance of REIDS is the existence of a range of species, each with a set of particular properties, such as wavelength, coherence etc. Combined with different hosts this results in a wealth of possibilities and tailored applications for different materials. Particularly, six-hour coherence time has been demonstrated in $Eu^{3+}:Y_{2}SiO_{5}$ by M. J. Zhong et al.\cite{zhong2015optically}. Erbium ions doped solids have attractive transition wavelengths in the telecom C-band, large inhomogeneous broadening up to THz\cite{xi2020experimental,saglamyurek2015efficient,veissier2016optical} and long coherence time over 1 second\cite{ranvcic2018coherence}. This would be a remarkable material for quantum memories and quantum network\cite{saglamyurek2015quantum,askarani2019storage,craiciu2019nanophotonic}, for which quantum frequency conversion is no more needed for long distance photon transmission. 

\section{Quantum network with other physical systems and hybrid approaches}
In addition to the above physical systems that can be used as quantum memory for implementation of quantum network, there are also some other systems under consideration such as quantum dots, warm atoms, mechanical oscillators and superconducting circuits, etc. This is highlighted by the experiment in 2001 when B. Julsgaard et al. generated entanglement between two warm atomic ensembles in the continuous variable domain\cite{julsgaard2001experimental}. In 2011, an entangled continuous-variable state, in the form of a two-mode squeezed state, was stored in two warm caesium atomic ensembles\cite{jensen2011quantum}. In 2018, J. P. Dou et al. prepared a quantum memory in Cs atomic ensembles at room temperature by utilizing DLCZ protocol\cite{dou2018broadband}. In 2020, H. Li et al. further generated entanglement between two warm atomic ensembles\cite{li2020heralding} based on Ref. \cite{julsgaard2001experimental}. In 2021, D. N. Du et al. demonstrated the connection of two warm atomic ensembles through HOM-interference of two photons, which are produced in two room-temperature quantum memories and transmitted over 158 km\cite{du2021elementary}. Although this experiment has not generated entanglement between two quantum memories, it is an important step to realize quantum networks over several hundred kilometers. In 2011, K. C. Lee et al. generated motional entanglement between vibrational states of two spatially separated, millimeter-sized diamonds at room temperature\cite{lee2011entangling}. In 2016, A. Delteil et al. used quantum dot as a quantum memory and generated heralded entanglement between spins of two quantum dots\cite{delteil2016generation}. A. Narla et al. reported the generation of loss-tolerant entanglement between distant superconducting qubits with concurrent measurements\cite{narla2016robust}. In 2018, R. Riedinger et al. demonstrated entanglement between two micromechanical oscillators across two chips separated by 20 cm\cite{riedinger2018remote} and C. F. Ockeloen-Korppi et al. realized stabilized entanglement of massive mechanical oscillators, where the moving bodies are two massive micromechanical oscillators, each composed of about $10^{12}$ atoms\cite{ockeloen2018stabilized}. P. Kurpiers et al. implemented deterministic state transfer and entanglement between two superconducting qubits fabricated on separated chips\cite{kurpiers2018deterministic}. 

\begin{figure}[t]
\centering
\includegraphics[width=\textwidth]{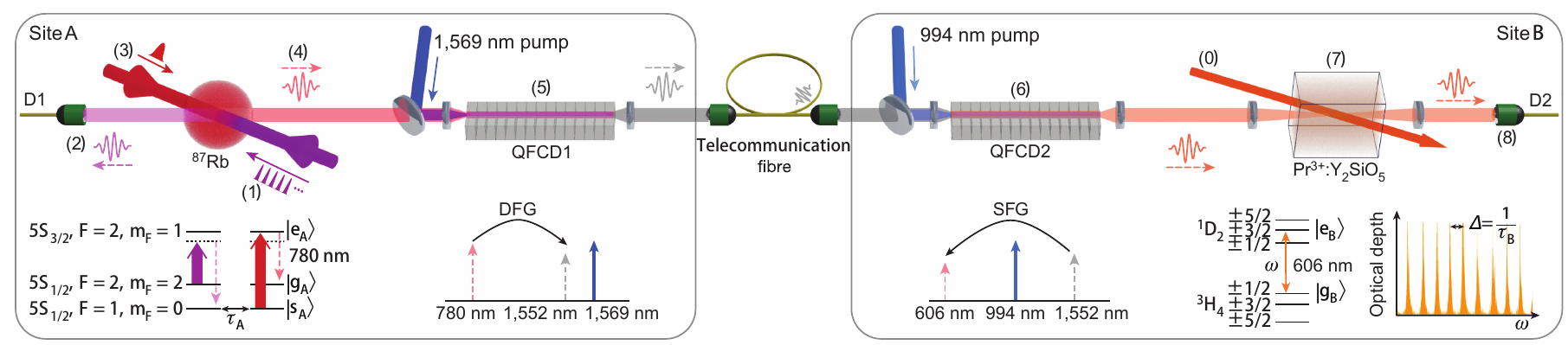}
\caption{Quantum state transfer from $^{87}Rb$ atomic ensembles to $Pr^{3+}:Y_{2}SiO_{5}$ crystal. In side A, the $^{87}Rb$ atoms is illustrated by write pulses (1) producing correlated photon pairs based on DLCZ protocol, and the write photons (2) herald the generation of spin-wave by D1. Then the read pulse (3) send to the $^{87}Rb$ atoms, and the spin-wave is read out, emitting the read-out photon (4). the read-out photons are converted from 780 nm to 1552 nm by QFCD1 (5). Then the converted photon is transmitted to side B through communication fiber. The photons are converted to 606 nm by QFCD2 (6), and finally store in the $Pr^{3+}:Y_{2}SiO_{5}$ crystal (7) ,which was prepared with a quantum memory by AFC protocol. After the read photons are recalled, they are ultimately detected by D2 (8). Inset: the energy levels of  $^{87}Rb$ atoms and $Pr^{3+}$ ions, a typical AFC structure. Adapted from Ref. \cite{maring2017photonic}.}
\label{fig:12}
\end{figure}

Different systems have different advantages and limitations for realizing quantum network, such as quantum information processing and storage. A hybrid quantum network could utilize the advantages of different systems. For instance, quantum gates can be operated easily in NV centers and single ions, CAEs can provide efficient and conveniently prepared entanglement source, while REIDS can provide sufficient capacity for storage of multi-mode photons and large bandwidth. For further quantum network, it would be useful to create quantum correlation between different species of quantum memories. In 2011, M. Lettner et al. demonstrated entanglement generation between two different systems (single atoms and Bose-Einstein condensate) separated by 30 meters\cite{lettner2011remote}. In 2017, N. Maring et al. connected two different quantum memories and demonstrated quantum state transfer between CAEs ($^{87}Rb$ atomic cloud) and REIDS ($Pr^{3+}:Y_{2}SiO_{5}$)\cite{maring2017photonic}, as shown in Figure \ref{fig:12}. 
In their experiment, first, they started to generate spin-wave inside $^{87}Rb$ atomic cloud in side A. The spin-wave was mapped onto a photonic time-bin qubit with a wavelength of 780 nm. Then the wavelength was converted from 780 nm to 1552 nm and transmitted to the side B through an optical fiber. On side B, the photons at 1552 nm were converted to 606 nm, and stored in the $Pr^{3+}:Y_{2}SiO_{5}$ crystal, which was prepared as a quantum memory using the AFC protocol. Finally the quantum state was transferred from $^{87}Rb$ atomic ensembles to $Pr^{3+}:Y_{2}SiO_{5}$ crystal. This experiment was an important step towards the realization of large-scale hybrid quantum networks. Except for utilizing all matter-based quantum memory as quantum node, all-optical quantum memory can also act as quantum node. In 2020, X. L. Pang et al. demonstrated room-temperature connection between atomic ensembles and an all-optical loop memory\cite{pang2020hybrid} by sending a write-out photon from the atoms to the optical loop memory.

\section{Quantum network performance estimates and other potential proposals}
The conventional quantum repeater scheme\cite{briegel1998quantum} has been widely researched and applied in various physical systems over the past two decades, it has been considered as one of the most efficient ways for constructing a global quantum network, however, there is still a long way to go since the entangled network greatly depends on the performance of quantum memories. In order to bloom everywhere, and then find a most practical way, there are still theoretical protocols continued to be proposed and analyzed towards a real-world quantum network. Below we will give a simple review on estimating common repeater performance and some other potential protocols, even though most of them are not demonstrated in experiments right now, they may provide unique solutions for construction of future quantum networks.
\subsection{Quantum repeater architecture and performance estimates}
{Similar with classical network, a developed quantum network inevitably has nodes with long distance between them and complex structure supporting advanced functions. Several theoretical investigations have been carried out to estimate the performance of existent quantum repeater protocol aiming at long-distance entanglement distribution and how to design new protocol to distributing entanglement in complex network topologies\cite{pirandola2019end,brito2020statistical,zhuang2021quantum}. \\
\indent
There have been a lot of protocols without considering error-correction proposed towards an origin quantum repeater, and several papers have been published on estimating the performance of the relevant protocols based on parameters of state of the art physical systems, such as entanglement sources, quantum memories and single photon detectors. According to the type of physical systems quantum memories used, the repeater protocols can be also categorized into the ones with single particle quantum memories\cite{santra2019quantum,krutyanskiy2019light,rozpkedek2019near} and with atomic-ensemble-based quantum memories\cite{sangouard2011quantum,brask2008memory,sinclair2014spectral,krovi2016practical,collins2007multiplexed,shchukin2019waiting,wu2020near}. \\
\indent
The single particle quantum memories have been used in theoretically studying the its performance for quantum repeater protocols like  trapped ions\cite{santra2019quantum,krutyanskiy2019light} and NV centers\cite{rozpkedek2019near}. Here, we take the NV centers for a detailed review. F. Rozpedek et al. gave four quantum repeater schemes and quantitively assessed their performance in creating entanglement distribution-based secret key when implementing them on setup based on NV centers in diamond and the typical parameters of this system in recent experiments or under expected cavity-enhanced configuration are adopted. The four schemes in this reference have been named as single sequential quantum repeater (SiSQuaRe) scheme, single photon scheme, single-photon with additional detection setup (SPADS) scheme, and single-photon over two links (SPOTL) scheme. The main differences among them are whether photon or electron spin is used at the end nodes, Bell state measurement (BSM) at the middle node is implemented for photons or for electron spin and $^{13}C$ nuclear spin, one or two middle nodes are set. When estimating the performance of these schemes, the authors used parameters including the dephasing and depolarizing parameters of $^{13}C$, photon-memory entanglement preparation time, and so on. Their calculations revealed that the secret-key rate achieved in single photon scheme can be more than seven times higher than the secret-key capacity, i.e., the theoretically maximum of secret-key rate in direct transmission via pure-loss channel. Moreover, The SPOTL scheme shows advantages in creating secret key at larger distance when the emitted photons from the NV centers are converted to those in telecom band and the fidelity of measurements are modestly improved.\\
\indent
For quantum repeater protocols with atomic-ensemble-based quantum memories, the imperfections of quantum memories have been considered in studying the performance\cite{brask2008memory,sinclair2014spectral,krovi2016practical,collins2007multiplexed,shchukin2019waiting}. Y. F. Wu et al.  evaluated the performance of four protocols with near-term experiment parameters\cite{wu2020near}. The four protocols have similar configurations in which photon sources and quantum memories are set at two end points of one elementary link while a beam splitter (BS) for photonic BSM is at the middle point. Another BS is set between the two elementary links for extending the entanglement via entanglement swapping. In the analysis, three key indices, the fidelity of the state after entanglement swapping, success probability of the entanglement swapping, and the average entanglement distribution time, are assessed. A significant progress of this work from previous ones is that the exponential decay of the efficiency of quantum memory is taken into consideration. The photon source used in numerical calculation includes quantum dots-based deterministic single or photon pair source, Rydberg atoms-based deterministic single or semideterministic photon pair source, and parametric down conversion-based photon-pair source, while the quantum memories includes Rare-earth-ion and Rydberg atoms-based ones. The numerical result shows that near-term technology is able to meaningfully realize the quantum repeaters protocols with relatively simple elements. The significant advancement over direct transmission, however, needs the further improvements. \\
\indent
A quantum version of classical network stack is proposed in Ref.\cite{dahlberg2019link}, in which the functions of the quantum physical, link, network and transport are allocated as attempting entanglement generation, robust entanglement generation, long distance entanglement and qubit transmission. The quantum nodes in the network can be classified into two types, namely, the controllable quantum nodes and automated quantum nodes. The authors focus their effort in designing a protocol for quantum link layer, which can supply a robust entanglement creation service to a quantum link connecting two controllable nodes. The desired services of this layer include requesting entanglement, response to entanglement requests, and fixed hardware parameters, while the throughput and latency are taken as the performance metrics which can be optimized in the implementation of this protocol. The proposed protocol is built up from five different components, i.e., distributed queue, quantum memory management (QMM), fidelity estimation unit (FEU), scheduler and the flow diagram of the operation. The protocol has been implemented on a discrete-event simulator, the data extracted from the quantum hardware system with NV centers are employed for validating the physical models in the simulation. The simulation results verify that protocol is robust even when classical control messages are significantly lost. In quantum network with multiple nodes, the entanglement routing is also important\cite{razaviintroduction,wengerowsky2018entanglement,pirandola2019end,pirandola2019bounds,joshi2020trusted}. The entanglement distribution can work in two modes, cintinuous and on-demond models, in which the entanglement routing problem reduces to routing entanglement on a virtual graph and a physical communication graph, respectively. K. Chakraborty et al. proposed three distributed routing algorithms in 2019\cite{chakraborty2019distributed}, i.e., modified greedy routing, local best effort routing and non-local best effort routing algorithms by borrowing ideas from complex network theory, and studied the performance of these algorithms on ring, grid, and recursively generated network topologies. The numerical results show that both of the modified greedy routing and local best effort routing algorithms perform better than the classical greedy routing algorithms in ring and grid topologies. When the quantum network gets complex, the bottleneck problems will probably occur, which means that many nodes requires quantum communication simultaneously and the quantum channel capacity may be exceeded. Similar with the solutions in classical network, quantum network coding is proposed by Hayashi et al. in 2007 to overcome this problem\cite{hayashi2007quantum}. The concepts in quantum repeater network also give drive force to quantum network coding technology. For instance, T. Satoh et al. have proposed a protocol that works in quantum repeater networks\cite{satoh2012quantum}, and T. Matsuo et al. further simplified this protocol by introducing measurement-based quantum computing\cite{matsuo2018analysis}.

}
\subsection{Next-generation quantum repeater and quantum network}
\begin{figure}[t]
\centering
\includegraphics[width=16cm]{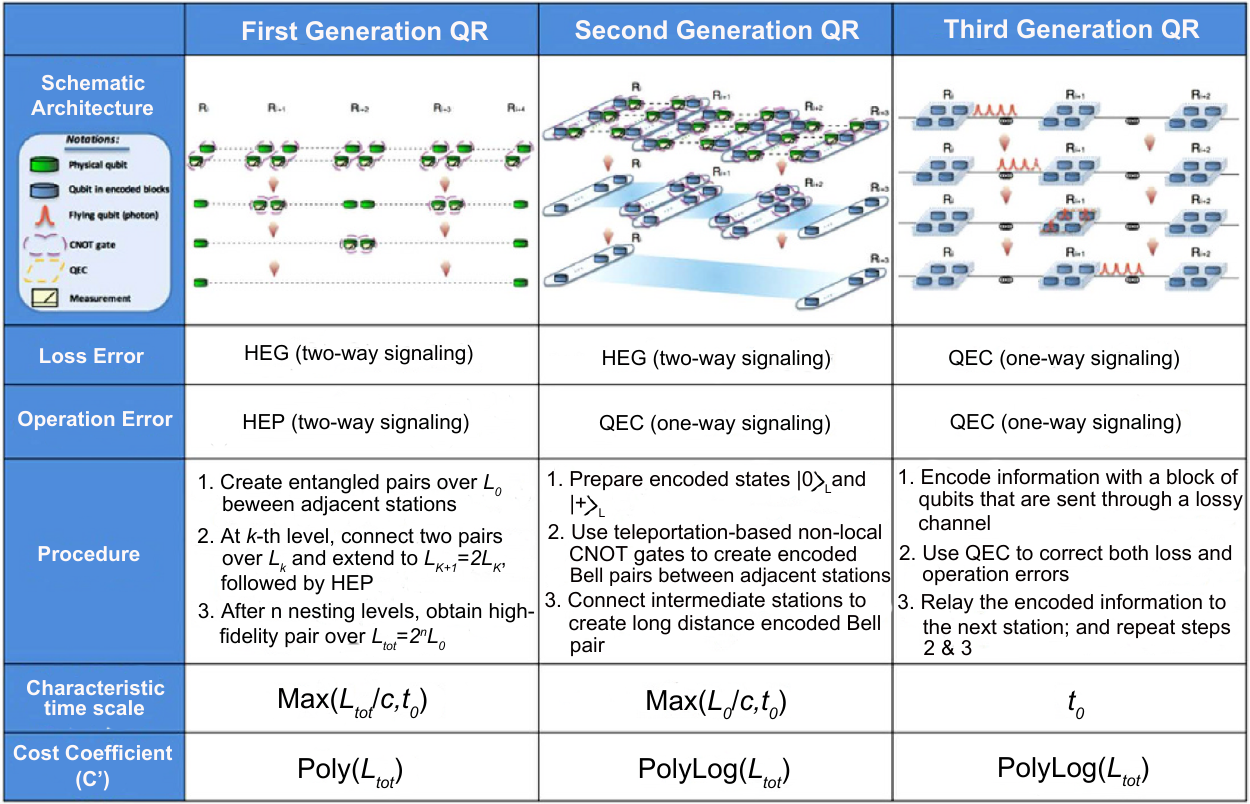}
\caption{Comparison of three generations of quantum repeaters. Adapted from Ref. \cite{muralidharan2016optimal}.}
\label{fig:13}
\end{figure}

Quantum repeaters can conquer both photon loss and operation errors, as a significant role for long distance quantum communication. However, the usually mentioned quantum repeater\citep{briegel1998quantum} employs both heralded entanglement and classical communication, which significantly limit the entanglement generation rate, specially, the time of classical communication for higher level entanglement swapping sets a minimal waiting time to next-step operation and simultaneously improves the requirement for memory lifetime. In order to get rid of the limitations, one-way quantum repeater is proposed, i.e., the third generation of quantum repeater mentioned below. In 2012, W. J. Munro et al. provided a repeater scheme capable using the redundant quantum parity code to achieve tolerate photon loss exceeding 50\% between different quantum nodes\cite{munro2012quantum}. The core of realization of this scheme was the use of parity code and the light-matter interaction for qubits transfer. It used the encoded matter qubits as logical qubits to operate the photon qubits for errors correction. More critically, the scheme did not require long-lived quantum memories, which was expected to tremendously increase the current quantum communication rate. As the proposal here mainly focused on the correction of photon loss errors using quantum parity code with an assumption of perfect gate operation, fault tolerance isn't under consideration in their analysis. In 2014, M. Sreraman et al. further proposed a fault-tolerant architecture, i.e., an ultrafast and fault-tolerant scheme for one-way quantum repeaters\cite{muralidharan2014ultrafast}. Like the scheme proposed by W. J. Munro et al., it also needs the matter qubits to manipulate and correct photons errors. Differently, they used a teleportation-based error correction (TEC) encoded procedure consisting of Bell state preparation and Bell measurement in each repeater node. For Bell state preparation, a (n,m)-quantum parity code is used, which is a class of Calderbank-Shor-Steane (CSS) codes for matter qubits. Further in 2016, in consideration of the existent repeater schemes and possible development trends of quantum repeaters, S. Muralidharan et al. categorized quantum repeaters into three generations\cite{muralidharan2016optimal}. They systematically compared the characteristics of three generations of quantum repeaters shown in Figure \ref{fig:13}. 
Different methods are adopted to correct loss and operation errors in three generations of quantum repeaters. The first generation of quantum repeaters respectively makes use of heralded entanglement generation (HEG) and heralded entanglement purification (HEP) to suppress loss and operation errors. Both of them require two-way classical signaling which set a limit for the communication rate. The second generation of quantum repeaters uses HEG for solve the problem of loss errors and quantum error correction (QEC) for operation errors where only one-way classical signaling is required. It is worth mentioning that the third generation of quantum repeaters correct both loss errors and operation errors via quantum error correction, the limitations by two-way classical signaling no longer needs to be in consideration, which can significantly benefit the achievement of high-speed quantum communication. \\
\indent
Another protocol to get off the limitations of quantum memories is to remove the quantum memories and build all-photonic quantum repeater with multi-photon entangled states, which was introduced by A. Koji et al. in 2015\cite{azuma2015all}. They proposed the all-photonics time-reversed quantum repeaters scheme without quantum memories as well, which did not need efficient coupling between photons and a matter qubit any more. In the scheme, it used a fast active feedforward technique for a loss-tolerant measurement and cluster-state flying qubits to replace Bell pairs, which can realize to send and receive the heralding signals for the entanglement swapping in the same repeater node. This scheme can make the rate of communication greatly increased and polynomially change with the channel distance. A proof of principle of such kind of all photonic quantum repeater with GHZ state without quantum memories was experimentally demonstrated by Z. D. Li et al. in 2019\cite{li2019experimental}. They experimentally realized the all-photon quantum repeater scheme of two parallel channels and one node with an overall fidelity of 0.606 $\pm$ 0.010. An 89\% enhancement of entanglement-generation rate over standard parallel entanglement swapping also proves the potential of all photonic quantum repeater compared to the repeater schemes with parallel Bell states and entanglement swapping.

\subsection{Quantum network architectures with satellites}
An alternative way to overcome the distance limitation is satellites based quantum network\cite{simon2017towards,mastriani2020satellite}, quantum communication with satellites has been also viewed as the most promising technology to distribute entanglement across large distances  \cite{bonato2009feasibility,bourgoin2013comprehensive,hosseinidehaj2018satellite,dequal2021feasibility,brito2021satellite} and several great experimental progresses has been achieved\cite{vallone2015experimental,takenaka2017satellite,yin2017satellite,liao2017satellite,ren2017ground,yin2017satellite,liao2018satellite}. Since the photon loss in outer space is much lower than that in atmosphere or fibers, extending quantum communication distance by satellite based quantum network seems a feasible way. The main difficulty occurs at the collection of photons because of the beam divergence over long distance transmission and movement of the satellites around the earth, as a solution, high-Earth-orbit satellites cover larger region and provide longer flyby time, which can efficiently improve the connection of satellite nodes and ground nodes. Another difficulty is the launch cost and fabrication technique for successfully placing satellites as quantum nodes. 

\begin{figure}[t]
\centering
\includegraphics[width=14cm]{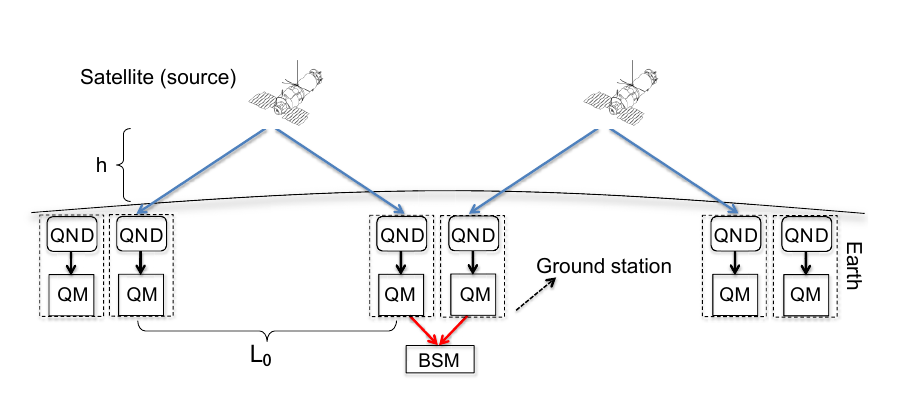}
\caption{Quantum network with satellites. Each elementary link with ground distance of $L_0$ is composed of an entangled source on a low-Earth-orbit satellite and two ground nodes, which consist of quantum nondemolition (QND) devices and quantum memories (QM). The successful distribution of entangled photons to each ground node is heralded by QND. Then in each node, the photons is stored by QM until entanglement between two QM in generated. Finally the entangled distance will be extended via Bell-state measurement on two elementary links. Adapted from Ref. \cite{boone2015entanglement}.}
\label{fig:14}
\end{figure}

Even there is a long way to go for improving the satellite performance and reducing the cost, constructing satellites based quantum network is believed as a once and for all project. In 2009, C. Bonato et al. discussed the feasibility and expected key generation rate of quantum key distribution between a LEO (low-Earth-orbit) satellite and a ground station\cite{bonato2009feasibility}. In 2015, K. Boone et al. discussed the approach of combining LEO satellites repeaters with entangled sources and ground stations together with quantum nondemolition measurements and quantum memories to make the quantum communication spread in a global range, as shown in Figure \ref{fig:14}\cite{boone2015entanglement}, 
where entanglement distribution rates were assessed with different satellite numbers and height over the distance of 20,000 km under a simple assumption that all satellites and stations are around the equator. In 2021, M. G{\"u}ndo{\u{g}}an et al. further presented a new approach which uses satellites with QM as repeater node to connect two ground stations\cite{gundougan2021proposal}. Different from Boone’s approach, here QMs are equipped on satellites rather than ground stations. The advantage is that the internode loss can be reduced due to that most of the involved nodes are launched into outer space, thus the weather can only influence few ground stations. According to their research, it’s reported that the distribution rates were three orders of magnitude faster than other global distance protocols through their architecture with existent protocols. Except for the presented proposals for satellite based quantum communication, there are also several detailed analyses in consideration of various influencing factors towards the realizations with these proposals. For instance, in 2013, J. P. Bourgion et al. discussed the design of LEO satellite quantum communication considering beam waist, wavelength, pointing error and telescope design, etc.\cite{bourgoin2013comprehensive}. In 2019, D. Vasylyev et al. discussed the influence of the ground observer's zenith angle, geographical latitude, and the meridian inclination angle of the satellite, etc.\cite{vasylyev2019satellite} and C. Liorni et al. analyzed the beam effects and weather dependence\cite{liorni2019satellite}. Further in 2021, C. Liorni et al. analyzed the performance of satellite based quantum communication by comparing entanglement distribution rate with different schemes such as scheme with ground sources and ground repeaters (GG), scheme with orbiting sources and ground repeaters (OG) and scheme with orbiting sources and orbiting repeaters (OO), in consideration of different number of nodes, distance and height of satellites\cite{liorni2021quantum}. According to the analysis, compared with other schemes, scheme with orbiting sources and orbiting repeaters shows the highest secret key rate over long distance. In the same year, S. Pirandola shows the ultimate limits and the practical rates of satellite based quantum communication in consideration of downlink/uplink configurations, various altitudes and zenith angles and night/day operations\cite{pirandola2021satellite}. Furthermore, he also considered the implementation of satellite based CV-QKD, which has also been presented the feasibility by D. Dequal. et al.\cite{dequal2021feasibility} and it may be suitable for high-rate quantum communication in the LEO and sub-LEO regions in the future. Based on the rich analyses mentioned above, satellite based quantum communication, is shown to be able to get a clear advantage over the optimal performance achievable by linear chains of quantum repeaters on the ground, and thus satellite based quantum communication is believed to be one of the most efficient ways to achieve a much higher quantum communication rate.  \\
\indent
Besides of the satellite based quantum communication that has been mostly analyzed, satellite based quantum network also presents great potentials in quantum sensing like time standards and frequency transfer, global navigation, high-precision gravimetry, etc. and also in tests of the core foundations of physics like gravitational wave observation\cite{kaltenbaek2021quantum}. To make the satellite based quantum network come true earlier, there are two main technical issues need to be improved before it become practical, i.e., performance of QM, cost and volume of equipment. Since satellite based quantum network has revealed its merits as a great candidate for global quantum communication, it must be the next research goal for a future quantum network.
 
\subsection{Quantum network with emerging new hardware approaches}
In most quantum repeater schemes, higher level entanglement swapping operations of matter-matter entanglement are utilized to build a long-distance quantum network between elementary links. However, the successful probability of entanglement swapping based on linear optics is relative low\cite{sangouard2011quantum}. A possible method to improve the successful probability is to use an auxiliary photon\cite{grice2011arbitrarily,wein2016efficiency}, which, however, increases the complexity and compounds errors.

To solve this problem, in 2018, F. Kimiaee Asadi et al. proposed a novel quantum repeater scheme to build long-distance quantum networks beyond the linear optics-based entanglement swapping\cite{asadi2018quantum}. In this scheme, each quantum node consists of two individual ions, i.e., $Er^{3+}$ and $Eu^{3+}$, both ions are placed in a microcavity. Similar with other repeater schemes, the first step is to realize light-matter entanglement. Here entanglement between the spin state of an $Er^{3+}$ and a spontaneous emitted photon are generated. Then though a joint Bell-state measurement on two spontaneous emitting photons from two different $Er^{3+}$, the maximally entangled state between spin states of two ions can be generated, forming a $Er^{3+}$- $Er^{3+}$ elementary link. Next, the quantum state of $Er^{3+}$ ions are mapped to $Eu^{3+}$ ions in the same microcavity by electric dipole moment and the entanglement between two $Er^{3+}$ ions is transferred to two $Eu^{3+}$ ions. After the mapping, the $Er^{3+}$ ions are re-initialized and generation process for a new $Er^{3+}$-$Er^{3+}$ elementary link can be repeated between the other neighboring $Er^{3+}$ ions. Finally, using entanglement swapping of electric dipole interaction on $Er^{3+}$ and $Eu^{3+}$ ions, the entanglement distance will be extended with high entanglement distribution rate. The scheme combines the advantages of $Er^{3+}$ and $Eu^{3+}$ ions, where $Er^{3+}$ can provide the interface of matter qubits and photonic qubits at telecom wavelength while $Eu^{3+}$ can offer long storage time due to the up to six-hour coherence lifetime\cite{zhong2015optically}. \\
\indent
As is well known that the key point for global quantum network is to achieve high quantum entanglement distribution rate, however, it is limited by heralded building entanglement, the storage lifetime of quantum memory and two-way signaling\cite{seri2017quantum,borregaard2019quantum}. To unleash the limitations, many protocols based on one-way repeaters and quantum error-correcting codes have been proposed\cite{fowler2010surface,munro2012quantum,azuma2015all}. In 2020, J. Borregaard et al. further proposed a new one-way quantum repeater architecture relying on photon-emitter interface\cite{borregaard2020one}. In this scheme, photonic tree-cluster states\cite{varnava2006loss} are utilized to encode quantum information and be transferred to next repeater node. Firstly signal qubit to be transmitted is encoded with the root spin qubit of a photonic tree cluster through a Bell measurement. Then the encoded signal qubit is sent to the distant next repeater node. At the next repeater node, the qubit is re-encoded between one of the first-level photon qubits and the root spin qubit of the new tree by Bell measurement. At the end node, the signal qubit can be transmitted to a receiving spin qubit by means of a controlled-phase gate. In this way, a quantum repeater network based on photonic tree-cluster states is built. Significantly, in this scheme, the multi-photon entangled state is generated, which are helpful to correct the photon loss before transmitting the signal qubit to the next node. In addition, O. Katz et al. also proposed a quantum memory protocol based on a new physical mechanism\cite{katz2020optical} in the same year, which is to map the photonic state to long-lifetime but optically inaccessible noble-gas spins collective state. This scheme provide a new hardware to realize light-matter entanglement and further build a global quantum network.

\section{Conclusion and discussion}
In this review we have given an account of the experimental and theoretical progress on constructing quantum networks. In recent years, great progress towards constructing quantum network has been made, however, a global quantum network is still far beyond the current state of the art. As discussed in Sec. 2 to Sec. 6, the developments of quantum networks with various physical systems and different theoretical schemes are at different stages. Some have successfully operated quantum gates; Some have demonstrated capacity with multi-mode storage, long coherence time, high efficiency and high fidelity of light-matter entanglement; Some have realized node-to-node entanglement; While some are still at an earlier stage of theoretical research. Table \ref{tab:1} summarizes the recent experiment progresses on entanglement of quantum nodes with different physical systems, and each system has its own advantages and limitations. 
\begin{table*}[t]
\centering
\caption{\bf\centering Overview of different systems to two node entanglement$^a$}
\begin{tabular}{ccccccc}
\hline
Physical system & $\eta_{g}$$^b$ & $\eta_{v}$ & overall efficiency$^c$ & $P_{atom-atom}$ & \makecell*[c]{Experimental \\ counting rate} & Fidelity \\
\hline
Single atoms \cite{ritter2012elementary} & 0.12 & 0.009 & $1\times10^{-3}$ & $3.2\times10^{-5}$ & 3 per min & 98 \% \\
Single atoms \cite{hofmann2012heralded}$^d$ & $1.08\times10^{-3}$ & 0.6 & $6.5\times10^{-4}$ & $0.19\times10^{-6}$ & 0.6 per min & 92 \%  \\
Single atoms \cite{rosenfeld2017event} & $1.25\times10^{-3}$ & $\geq0.98$ & $1.23\times10^{-3}$ & $0.69\times10^{-6}$ & 2.2 per min & -\\
Atomic ensembles \cite{yuan2008experimental} & $2.5\times10^{-3}$ & 0.15 &  $3.75\times10^{-4}$ & $7\times10^{-8}$ & 0.04 per min & 83 \% \\
Atomic ensembles \cite{jing2019entanglement} & $2\times10^{-2}$ & 0.3 & $6\times10^{-3}$ & $1.8\times10^{-5}$ & 32.4 per min & 85 \%  \\
Atomic ensembles \cite{yu2020entanglement} & $\sim 4\times10^{-3}$ & 0.26 & $\sim 1\times10^{-3}$ & $0.73\times10^{-6}$ & 0.4 per min & 72 \% \\
Trapped ions \cite{moehring2007entanglement} & $1.2\times10^{-4}$ & $\leq1$ & $\leq1.2\times10^{-4}$ & $3.6\times10^{-9}$ & 0.12 per min & 63 \% \\
Trapped ions \cite{olmschenk2009quantum} & $2.8\times10^{-4}$ & $\leq1$ & $\leq2.8\times10^{-4}$ & $2.2\times10^{-8}$ & 0.083 per min & 90 \%  \\
Trapped ions \cite{maunz2009heralded} & $2.85\times10^{-4}$ & $\leq1$ & $\leq2.85\times10^{-4}$ & $4.2\times10^{-8}$ & 0.09 per min & 89 \%  \\
Trapped ions \cite{hucul2015modular} & $4.4\times10^{-3}$ & $\leq1$ & $\leq4.4\times10^{-3}$ & $9.7\times10^{-6}$ & 27 per min  & 78 \%   \\
Trapped ions \cite{stephenson2020high} & $2.1\times10^{-2}$ & $\leq1$ & $\leq2.1\times10^{-2}$ & $2.18\times10^{-4}$ & 182 per sec  & 94 \%  \\
NV centers \cite{bernien2013heralded} & $4.5\times10^{-4}$ & $\leq1$ & $\leq4.5\times10^{-4}$ & $1\times10^{-7}$ & 0.1 per min & 73 \%  \\
NV centers \cite{hensen2015loophole} & $2\times10^{-4}$ & $\leq1$ & $\leq2\times10^{-4}$ & $6.4\times10^{-9}$ & 0.017 per min & 92 \%  \\
REIDs \cite{Puigibert2019a} & - & - & - & $3.14\times10^{-8}$ & 2.5 per sec & 93 \% \\
REIDs \cite{liu2021experimental} & - & - & - & $7.5\times10^{-12}$ & 1.1 per hour & 80 \% \\
Quantum dots \cite{delteil2016generation} & $1.4\times10^{-4}$ & 0.002 & $2.8\times10^{-7}$ & $3.9\times10^{-14}$ & 0.003 per hour & 55 \%  \\

\hline
\end{tabular}
\begin{flushleft}
\footnotesize {$^a$ In this table, the matter-matter entanglement is through single photon detection, i.e. Fock-state entanglement are not involved.  }

\footnotesize {$^{b} \eta_{g}$ is the overall generation efficiency of light-matter entanglement, containing the overall detected efficiency of photon. $\eta_{v}$ is the verification efficiency of the matter qubits. Overall efficiency is the product of $\eta_{g}$ and $\eta_{v}$. $P_{atom-atom}$ is the total successful probability of generating and verifying entanglement of two quantum memories per experimental trial. Experimental counting rate is the number of detected coincident events with entanglement generation and verification in unit time. Fidelity is the measured fidelity of two-node entangled state. }

\footnotesize {$^c$ In this table, the $\eta_{v}$ of trapped ions and NV centers can reach unity with fluorescence detection. Other systems like single atoms, CAEs are mostly based on single photon detection, which has a relatively low converting and detection efficiency. It is worth noting that in Ref. \cite{rosenfeld2017event}, the $\eta_{v}$ is also nearly one with photoionization detection.}

\footnotesize {$^d$ In some References, two pairs of light-matter entanglement are produced with different efficiency. Here, we review the average of the two pair of light-matter entanglement \cite{hofmann2012heralded,rosenfeld2017event}. The value of $\eta_{g}$ in Ref. \cite{yu2020entanglement} is estimated due to the read-out efficiency and $P_{atom-atom}$. In Ref. \cite{jing2019entanglement}, the three quantum nodes are entangled. Here, the efficiency of two nodes of entanglement generation and verification and counts rate are presented.}
\end{flushleft}
  \label{tab:1}
\end{table*}

At present, although many theoretical schemes in principle allows the distance extension and the number increase of quantum nodes, it remains experimentally difficult to have practical applications. The main reason is that finite experimental conditions limit the entanglement generation rate and fidelity, thus, to extend the distance and number of nodes, further improvement of the experimental technique is needed. 

The first challenge is to reduce the loss of photon transmission. There are mainly two ways to solve it: one is to establish free-space quantum channel via satellites\cite{peng2005experimental}; the other one is to build fiber-based quantum channels. Due to the lower optical loss caused by absorption in vacuum than in atmosphere, extending photon communication distance by satellite repeaters is a feasible method. In particular, successful experiments of long-distance quantum key distribution\cite{liao2017satellite} and quantum teleportation\cite{ren2017ground} via ‘Micius’ satellite open a new way for building satellite-based quantum entanglement networks. Besides, due to photons at the wavelength around 1.5 $\mu$m transmitting in the optical fibers with low loss (typically about 0.2 dB/km), telecom-band photons can be also utilized to connect long-distance quantum nodes through fibers. However, in most experiments on photon-atom entanglement, the wavelength of photons is far from telecom wavelength. It is difficult to directly use these photons in optical fibers. One method to solve this problem is to utilize efficient quantum frequency conversion\cite{kumar1990quantum}, with which the fiber distance of two quantum nodes has been extended over 50 km\cite{yu2020entanglement}. Another method is to employ wavelength-bridging entanglement or use atoms with suitable transition wavelength (e.g., erbium atoms). In 2021, two REIDS are entangled directly using telecom-band photons that are generated by wavelength-bridging entanglement\cite{lago2021telecom}. Although the entangled fiber distance is only 50 m, it presents the potential to realize long-distance fiber-based quantum network with multimode quantum memories. Furthermore, $Er^{3+}$ ions-doped solids have attractive transition wavelength in the telecom band, which are able to directly allow quantum storage of telecom-band photons. Although the entanglement between two $Er^{3+}$ ions-doped solids have not been generated, the remarkable material is one of the most promising materials to construct long-distance quantum network. In the future, we believe that quantum network is fiber-based and free space-based hybrid networks just like the existent classical communication.

The second challenge is to realize a high-performance light-matter interface. To achieve this goal, quantum memory with high efficiency, long storage time, high fidelity, wide bandwidth, multimode capacity is required. At present, quantum memories with high storage fidelity of 0.999\cite{zhou2012realization} or large multimode capacity of 1650 modes\cite{wei2021multiplexed} have been demonstrated, storage time exceeding one second can also in principle be realized\cite{longdell2005stopped}, etc. However, so far there are still no quantum memories which can simultaneously have all advantages over these parameters. The difficulties that need to be overcome vary in different quantum memories. In the future, it will be appealing to investigate new kinds of quantum memories as quantum nodes or combining the different quantum memories together to construct a hybrid platform based quantum network.

The third challenge is to improve the efficiency of matter-matter entanglement. In addition to improving the efficiency of light-matter entanglement, the high probability and fidelity of entanglement swapping is also required. At present, most of entanglement swapping is not deterministic, leading to low success probability for experimental implementation. There are some methods to solve this problem. For instance, deterministic local two-qubit gates are achieved using the Coulomb interaction of trapped ions in trapped ions system\cite{blatt2008entangled,sangouard2011quantum} or the dipole interaction of atoms excited to a Rydberg state in neutral atomic system\cite{saffman2010quantum}. Besides, a mutual electric dipole-dipole interaction between nearby ions trapped in the same cavity was proposed to implement entanglement swapping\cite{asadi2018quantum}. In the future, more simple and accessible deterministic entanglement swapping operations are needed to further improve the experiments conditions.

The fourth challenge is to increase the repetition frequency of experiment. To increase the frequency of experimental repetition, many key experimental technologies need to be further improved, especially bandwidth. We take broadband quantum memory as an example, to achieve the experimental repetition frequency of GHz, the temporal width of stored photons should be as short as sub-ns level, which requires the storage bandwidth of quantum memory up to GHz. However, it is difficult to achieve in many EIT/DLCZ based physical systems, such as single atoms, trapped ions, NV centers, etc. Remarkably, except for Raman quantum memory, REIDS with large inhomogeneous broadening up to hundreds of GHz have become candidates for practical high-speed and multiplexed quantum memory. Currently, a quantum memory with a single memory bandwidth of 10 GHz and total available bandwidth up to 50 GHz has been implemented in REIDS\cite{wei2021multiplexed} to store sub-ns photonic wavepacket. REIDS is a powerful physical system to achieve long-distance high-speed quantum networks.

In general, different systems have different advantages. In NV centers and single atoms, quantum gates can be operated easily. In CAES, light-matter entanglement can be prepared with high-efficiency and convenience. While in REIDS, it can provide suﬀicient capacity for storage of multi-mode photons and large bandwidth. Although there is still a long way to go due to the aforementioned challenges, combining the advantages of different physical systems to build a hybrid quantum network is one of the promising paths. In addition to those already implemented quantum networks in real experiments, there are still some promising theoretical protocols proposed to construct real-world quantum networks, such as one-way quantum repeater networks, satellites based quantum networks, and quantum networks with new hardware approaches. If some key experimental technologies are addressed, these schemes will further drive the development of future quantum networks.

For deeper understanding of quantum networks with specific quantum memories, we would recommend readers pay attention to some other excellent reviews on quantum memories and quantum networks\cite{wehner2018quantum,lvovsky2009optical,simon2010quantum,heshami2016quantum,reiserer2015cavity,sangouard2011quantum,duan2010colloquium,awschalom2018quantum}.


\setlength{\parindent}{0em}
\medskip

\medskip
\textbf{Acknowledgements} \par 
This work was supported by the National Key Research and Development Program of China (Nos. 2018YFA0307400, 2018YFA0306102), National Science Foundation of China (Nos. 61775025, 91836102, U19A2076, 62005039, 12004068), China Postdoctoral Science Foundation (Nos. 2020M673178, 2020M683275, 2021T140093).\par

\medskip
\textbf{Disclosures}. 
The authors declare no conflicts of interest.
%
\bibliographystyle{MSP}

\bibliography{Ref}




\clearpage
\thispagestyle{fancy}
\fancyhead{}
\textbf{Biography}
\begin{figure}[htbp]
\centering
\includegraphics[width=40mm]{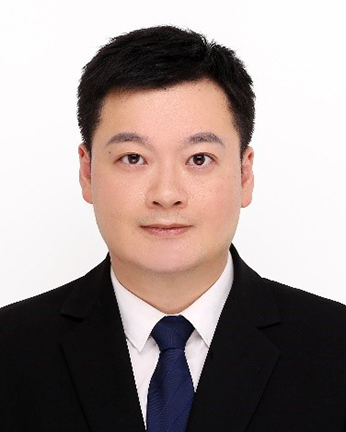}
\caption*{Qiang Zhou, born in 1984, studied Optoelectronic Information Science and Technology at the University of Electronic Science and Technology of China (China). He received his Ph. D. on fiber integrated quantum light sources in 2011 from Tsinghua University (China). He worked at the University of Calgary (Canada) as postdoctoral fellow from 2014 to 2017, where he has investigated quantum teleportation, quantum memory, quantum light source, measurement-device-independent quantum key distribution etc. He joined the University of Electronic Science and Technology of China (China) in 2017 as an assistant professor and obtained his full professor position in 2019. Dr. Zhou currently is a principal investigator for National Key R\&D Program of China. His main research interests focus on developing hardware for quantum Internet.}
\label{fig：111}
\end{figure}

\clearpage

\textbf{Table of contents}
\begin{figure}[htbp]
\centering
\includegraphics[width=55mm]{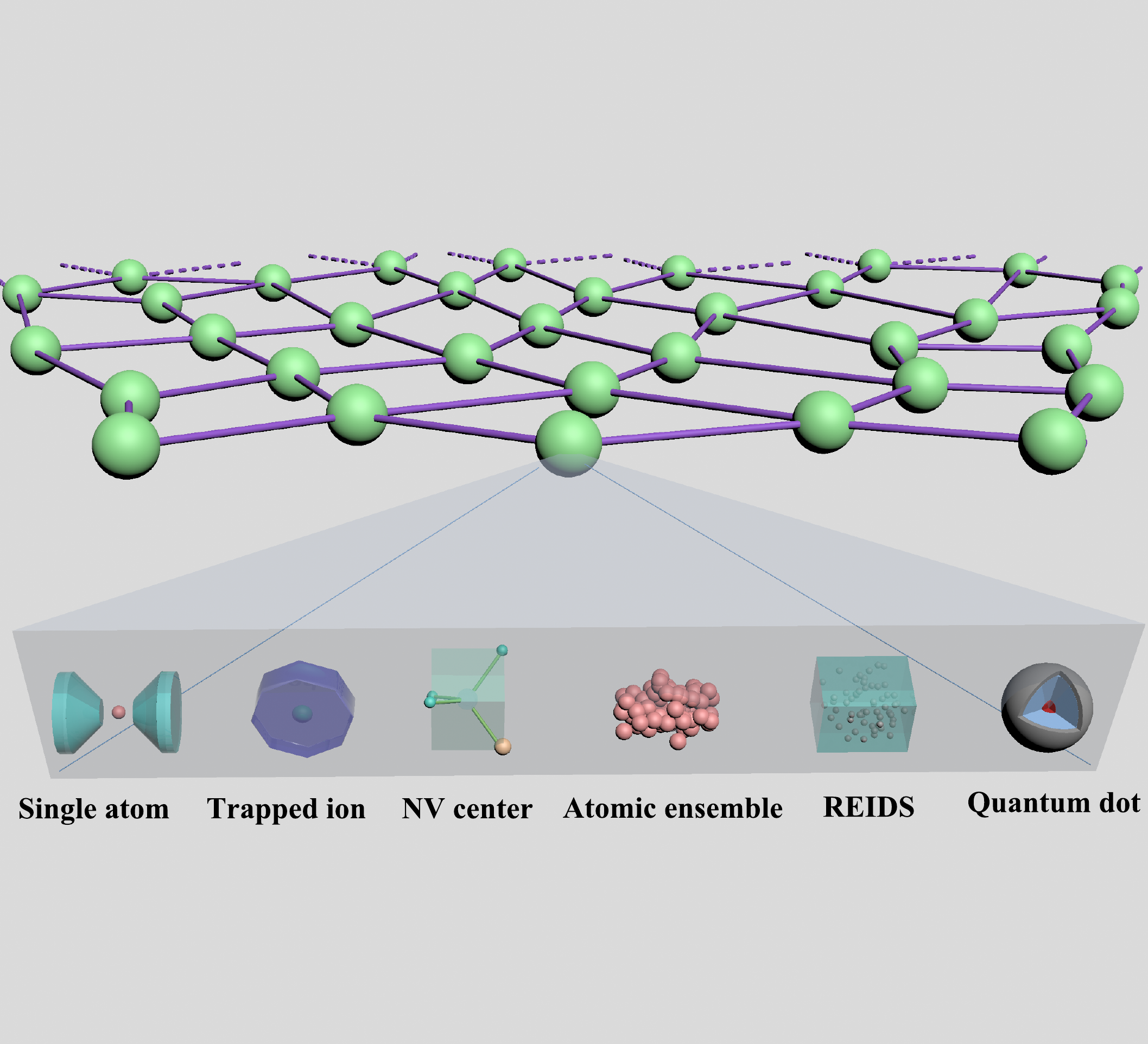}
\caption*{Quantum networks linking multiple remote quantum nodes consist of quantum memories served as stationary quantum nodes and flying photonic qubits served as quantum channels. This review summarizes and discusses the state of the art and future challenges for constructing quantum networks in various physical systems like single neutral atoms, cold atomic ensembles, trapped ions, NV centers, rare-earth-ions doped solids, etc.}
\end{figure}

\end{document}